\documentclass[fleqn,usenatbib]{mnras}

\usepackage{times}
\usepackage[T1]{fontenc}

\DeclareRobustCommand{\VAN}[3]{#2}
\let\VANthebibliography\thebibliography
\def\thebibliography{\DeclareRobustCommand{\VAN}[3]{##3}\VANthebibliography}

\usepackage{graphicx}	
\usepackage{amsmath}	
\usepackage{amssymb}	

\usepackage{rotating}
\usepackage{pdflscape}
\usepackage{colortbl}
\usepackage[normalem]{ulem}
\usepackage{color}
\usepackage{bm} 
\usepackage{orcidlink}
\definecolor{darkblue}{rgb}{0.0,0.0,0.8}
\definecolor{darkred}{rgb}{0.75,0.,0.25}
\definecolor{darkorange}{rgb}{1,0.3,0.0}
\definecolor{darkgreen}{rgb}{0.0,0.6,0.0}
\definecolor{darkpurple}{rgb}{0.8,0.,0.9}
\definecolor{brown}{rgb}{0.65,.16,0.16}
\definecolor{grey}{rgb}{0.4,0.5,0.6}
\definecolor{white}{rgb}{1,1,1}
\definecolor{trolleygrey}{rgb}{0.5, 0.5, 0.5}
\definecolor{lavender}{rgb}{0.835,0.812,0.969}
\definecolor{pastelorange}{rgb}{0.99,0.92,0.82}
\definecolor{pastelblue}{rgb}{0.85,0.93,0.99}

\newcommand{\darkg}[1]{\textcolor{trolleygrey}{#1}}

\newcommand{\darkor}[1]{\textcolor{darkorange}{#1}}

\newcommand{\mpo}{{\sc MAMPOSSt-PM}}
\newcommand{\balrogo}{{\sc BALRoGO}}
\newcommand{\agama}{{\sc Agama}}
\newcommand{\gaia}{{\sl Gaia}}
\newcommand{\hst}{{\sl HST}}
\newcommand{\cmc}{{\tt CMC}}
\newcommand{\parsec}{{\sc Parsec}}

\newcommand{\msun}{\rm M_\odot}
\newcommand{\kms}{\rm km\,s^{-1}}
\newcommand{\masyr}{\rm mas\,yr^{-1}}

\title[Dark central mass in M4]{An elusive dark central mass in the globular cluster M4}

\author[Vitral, Libralato, Kremer, Mamon, Bellini, Bedin \& Anderson.]{
Eduardo Vitral$^{\orcidlink{0000-0002-2732-9717}}$,$^{1,2}$\thanks{E-mail: evitral@stsci.edu} 
Mattia Libralato$^{\orcidlink{0000-0001-9673-7397}}$,$^{3}$\thanks{E-mail: libra@stsci.edu} 
Kyle Kremer$^{\orcidlink{0000-0002-4086-3180}}$,$^{4,5}$ 
Gary A. Mamon$^{\orcidlink{0000-0001-8956-5953}}$,$^{2}$ 
Andrea Bellini$^{\orcidlink{0000-0003-3858-637X}}$,$^{1}$
\newauthor
\, Luigi R.\ Bedin$^{\orcidlink{0000-0003-4080-6466}}$$^{6}$
and Jay Anderson$^{\orcidlink{0000-0003-2861-3995}}$$^{1}$
\\
$^{1}$Space Telescope Science Institute, 3700 San Martin Drive, Baltimore, MD 21218, USA \\
$^{2}$Sorbonne Universit\'e, CNRS, UMR 7095, Institut d’Astrophysique de Paris, 98 bis bd Arago, 75014 Paris, France \\
$^{3}$AURA for the European Space Agency (ESA), ESA Office, Space Telescope Science Institute, 3700 San Martin Drive, Baltimore, MD 21218, USA \\
$^{4}$TAPIR, California Institute of Technology, Pasadena, CA 91125, USA\\
$^{5}$The Observatories of the Carnegie Institution for Science, Pasadena, CA 91101, USA\\
$^{6}$Istituto Nazionale di Astrofisica, Osservatorio Astronomico di Padova, Vicolo dell'Osservatorio 5, Padova I-35122, Italy
}

\date{Accepted XXX. Received YYY; in original form ZZZ}

\pubyear{2015}

\begin{document}
\label{firstpage}
\pagerange{\pageref{firstpage}--\pageref{lastpage}}
\maketitle

\begin{abstract}
Recent studies of nearby globular clusters have discovered excess dark mass in their cores, apparently in an extended distribution, and simulations indicate that this mass is composed mostly of white dwarfs (respectively stellar-mass black holes) in clusters that are core-collapsed (respectively with a flatter core).
We perform mass-anisotropy modelling of the closest globular cluster, M4,
with intermediate slope for the inner stellar density.  
We use proper-motion data from \gaia\ EDR3 and from observations by the \textit{Hubble Space Telescope}. 
We extract the mass profile employing Bayesian Jeans modelling, and check our fits with realistic mock data.
Our analyses return isotropic motions in the cluster core and tangential motions ($\beta\approx -0.4$$\pm$$0.1$) in the outskirts.
We also robustly measure a dark central mass of roughly $800\pm300 \, \msun$, but it is not possible to distinguish between a point-like source, such as an intermediate-mass black hole (IMBH), or a dark population of stellar remnants of extent $\approx 0.016\,\rm pc \simeq 3300\,AU$. However, when removing a
high-velocity star from the cluster centre, the same mass excess is found, but more extended ($\sim 0.034\, \rm{pc} \approx 7000\,\rm AU$). We use Monte Carlo $N$-body models of M4 to interpret the second outcome, and find that our excess mass is not sufficiently extended to be confidently associated with a dark population of remnants.  Finally, we discuss the feasibility of these two scenarios (i.e., IMBH vs. remnants), and propose new observations that could help to better grasp the complex dynamics in M4's core.
\end{abstract}




\section{Introduction}

Few stellar systems in the Universe are as active and dynamically complex as globular star clusters (GCs). Indeed, the interplay between stellar evolution and dynamical interactions allows GCs to serve as laboratories for a vast number of interesting astrophysical phenomena, such as formation of black-hole mergers with components in the proposed pair-instability mass gap \citep[e.g.,][]{Rodriguez2019,DiCarlo2020,Kremer2020,GerosaFishbach2021}, gravitational waves \citep{LIGO1,LVC2021,Rodriguez+21b}, formation of compact black hole--luminous star binaries \citep[e.g.,][]{Strader2012,Giesers+19,Kremer2018xrb}, stellar-mass tidal disruption events \citep[e.g.,][]{Perets2016,Kremer2019b}, Type Ia supernovae \citep[e.g.,][]{Webbink1984}, formation of young neutron stars \citep[e.g.,][]{NomotoIben1985} and fast radio bursts \citep{Bhardwaj2021,Kirsten2021,Kremer2021frb,Lu2021}. Finally, 
one of the potential outcomes of these dense environments is intermediate-mass black holes (IMBHs, \citealt{Madau&Rees01,Miller&Hamilton02,PortegiesZwart&McMillan02,PortegiesZwart+04,Giersz+15,Gonzalez+21}, with masses $\sim 10^2 - 10^5 \, \msun$), thought to be the missing link of black hole evolution, with barely a few observed cases \citep[e.g.,][]{Chilingarian+18,Lin+20,TheLIGOScientificCollaboration+20}. Hence, this class is much in contrast with the many stellar-mass black holes ($\lesssim 10^2 \, \msun$) and supermassive black holes ($\gtrsim 10^5 \, \msun$), which have already been confirmed for a considerable amount of time \citep[e.g.,][]{Webster&Murdin&Murdin72,Bolton72,Hoyle&Fowler63,Schmidt63,EventHorizonTelescopeCollaboration+19}.

For this reason, many studies have targeted GCs to search for IMBH candidates (e.g., \citealt*{Gebhardt+02}; \citealt{Baumgardt+03}; \citealt*{Noyola+08}; \citealt{vanderMarel&Anderson&Anderson10,Baumgardt17,Kamann+16,Tremou+18,Haberle+21}), searching for electromagnetic signatures associated with accretion of material onto the IMBH and/or dynamical signatures (i.e., the dynamical effect of the IMBH on cluster stars). However, both of these detection methods face challenges. Accretion-signature searches are limited, since all IMBHs may not necessarily be actively accreting, hence not emitting light at observable frequencies. Additionally, some proposed IMBH accretion candidates may be more naturally explained as neutron stars accreting at super-Eddington rates \citep[e.g.,][]{Bachetti2014,RodriguezCastillo2020}. In the case of dynamical-signature searches, until recently the necessary completeness of the astrometric-quality data to detect IMBHs through the cluster stellar kinematics was not attained.  
Furthermore, proposed IMBH dynamical signatures may instead be explained by the presence of a sub-clustered population of faint stellar remnants in the centre of GCs \citep[e.g.,][]{Zocchi+19,Mann+19,Vitral&Mamon21,Vitral+22}.

Indeed,  given their higher masses, stellar remnants tend to naturally concentrate in the inner regions of GCs through mass segregation due to dynamical friction \citep{Chandrasekhar43}. First, the most massive remnants (i.e., black holes) sink to the cluster's centre and form a compact population that can
delay cluster core-collapse \citep{Henon61,LyndenBell&Wood&Wood68}, by means of \textit{black hole binary burning} (i.e., black hole-binary-mediated encounters that input energy into the inner regions of the GC, as explained in \citealt{Kremer+20p}). These black holes tend to be ejected however, mainly by means of dynamical interactions with other black holes \citep[e.g.,][]{Kremer+20} and natal kicks \citep[e.g.,][]{Repetto+12,Mandel16}, but some BHs also merge \citep[e.g.,][]{Rodriguez+21b}.  All of these factors result, eventually, in a negligible black-hole population, on $\gtrsim10\,$Gyr timescales.

Once this happens, other luminous stellar components will sink to the cluster cores, in addition to less massive compact objects such as neutron stars and the more massive white dwarfs. When these more luminous components collapse in the centre, forming the characteristic core-collapse inner cusp, \textit{stellar} and \textit{white dwarf} binary-burning effectively halts further shrinking of the core \citep{Kremer+21}. These populations of white dwarfs and black holes in the cores of core-collapsed and non core-collapsed GCs, respectively, tend to form a sub-cluster of roughly $0.1$~pc in size that can easily mimic an IMBH dynamical signature, if there are not enough tracers in the central region. Recently, in \citeauthor{Vitral+22} (\citeyear{Vitral+22}, hereafter Paper~I), we confirmed this by analysing two GCs: NGC~3201 (non core-collapsed) and NGC~6397 (core-collapsed) and assigned an extended population of black holes in the former and white dwarfs in the latter, both amounting up to roughly $1000 \, \msun$. 

In Paper~I, we used state-of-the-art proper motion data from observations from the \textit{Hubble Space Telescope}\footnote{The \hst\ data archive is available at \url{https://archive.stsci.edu/hlsp/hacks}.} (\hst, \citealt{Libralato+22}) and from \gaia\ EDR3 \citep{GaiaCollaboration+21}, which was fitted with the Bayesian Jeans mass-orbit modelling code \mpo\ (see \citealt{Mamon+13}; Mamon \& Vitral in prep. and \citealt{Read+21} for a comparison with other methods). We then compared our fits to outcomes from mock datasets constructed with \agama\ \citep{Vasiliev19a} and interpreted them with the help of Monte Carlo $N$-body models from \cmc\ \citep{Kremer+20}. Such high-precision data was complete enough to trace down very small extensions such as $\sim 0.1$~pc and rule out the possibility of an IMBH in those clusters. 
 
Additionally, the discovery of a ($\lesssim 100-500\,\msun$) IMBH in an old GC (such as most of those in the Milky Way), would pose an interesting problem, since such a black hole should probably have merged with other GC black holes during its lifetime, and hence would have been ejected due to huge gravitational wave recoil kicks \citep[e.g,][]{Peres62,Lousto+10,HolleyBockelmann+08}, usually far above the cluster's escape velocity \citep[e.g,][]{Merritt+04_recoilkick,Campanelli+07}.
On the other hand, massive IMBHs ($\gtrsim 1000\,\msun$) would be involved in high mass-ratio mergers and suffer less potent recoil kicks, under the escape velocity.\footnote{These massive IMBHs would be pushed out of the GC centre, without necessarily being ejected from the GC.}

In the current study, we set out to study the closest GC to our Sun, Messier 4 (hereafter M4, also known as NGC~6121), with the same data and methods recently validated in Paper~I.   M4's proximity and its many observations make it a very interesting source, worthy of special attention. Although this study is among the many to analyse the clustering of compact objects in M4 \citep[e.g.,][]{Richer+95,Bassa+04,Bedin+13,HenaultBrunet+19}, it is to our knowledge the first to provide constraints on the mass and extent of such a central sub-cluster, while simultaneously testing the possibility of an IMBH, by means of Jeans mass-orbit modelling, and state-of-the-art proper-motion data. 

%
We divide our work in the following manner: Section~\ref{sec: methods} briefly explains the data and methods we used, nearly identical to those presented in Paper~I; Section~\ref{sec: results} presents our main results and evaluates their robustness; Section~\ref{sec: discussion} further considers the reliability of our main results, interms of their feasibility and implications. Finally, Section~\ref{sec: conclusion} summarises and concludes the analysis.

\section{Data \& Methods} \label{sec: methods}

\begin{figure}
\centering
\includegraphics[width=0.95\hsize]{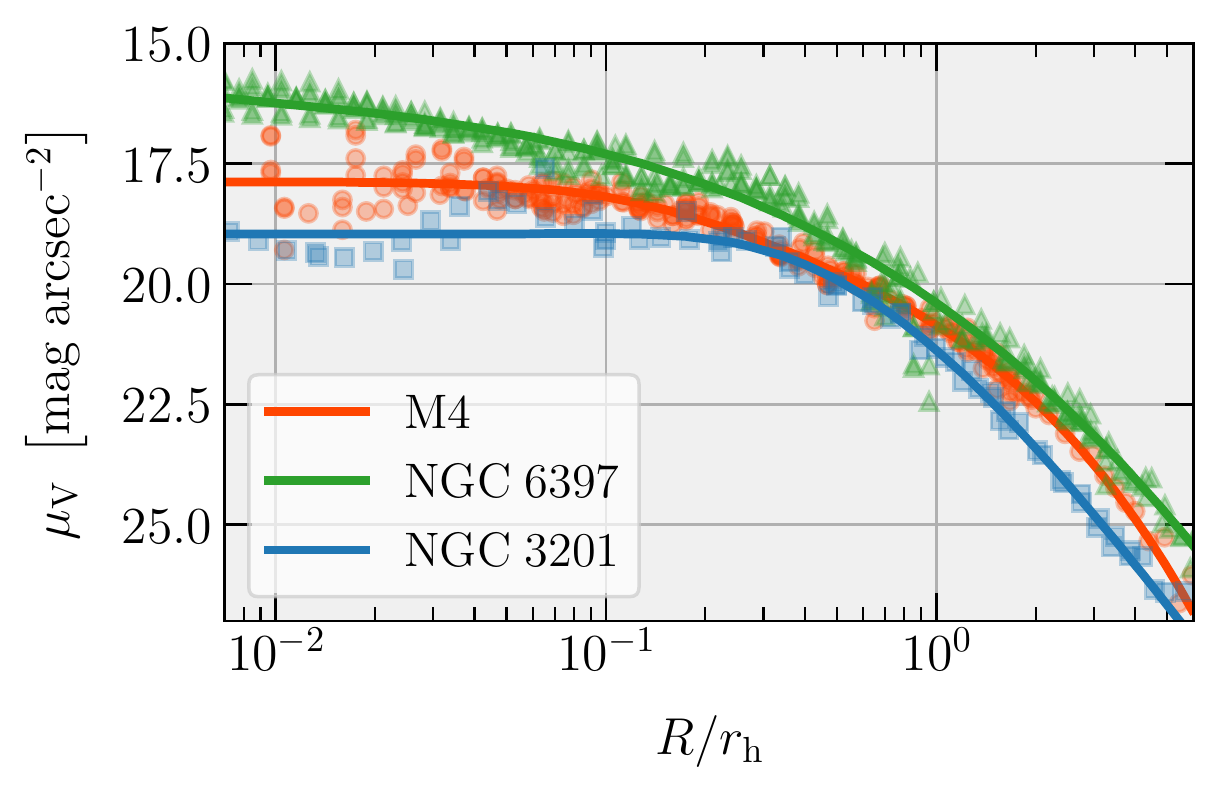}
\caption{\textit{Surface brightness profiles:}
Comparison of the surface brightness profiles as a function of projected radius $R$ (normalised by the half-light radius from \protect\citealt{Harris10}) for NGC~3201 (non core-collapse, \textit{blue squares}), NGC~6397 (post core-collapse, \textit{green triangles}) and M4 (non core-collapse, \textit{orange circles}), from \protect\cite{Trager+95}. The transparent symbols indicate the observed surface brightness profiles, while the solid lines indicate the respective Chebyshev fit.}
\label{fig: cc-comp}
\end{figure}

\subsection{M4 overview} \label{ssec: overview-m4}

The closeness of M4 makes its study not only particularly interesting, but very straightforward (when compared to much farther clusters).  Indeed, its proximity allows a higher fraction of stars to be measured with high-precision proper motions, with errors often smaller than the GC velocity dispertion by an order of magnitude. This in turn provides a sufficient number of inner stars (e.g., $\sim 1$ arcsec from its centre) to probe a possible inner dark mass.  Our main dataset after the cleaning routine explained in Section~\ref{ssec: cleaning} leaves us with 6158 \gaia\ EDR3 stars, and 4365 \hst\ stars, providing complete coverage from the cluster's interior\footnote{The \hst\ data we use extends from $0\farcs9$ up to $149\farcs2$.} out to its outermost radii.\footnote{\label{ftn:Re} The maximum allowed projected radius is set at $2\,R_{\rm e}$ (see Figure~\ref{fig: pos-m4}), as in Paper~I, where  $R_{\rm e}$ is the \emph{effective radius} containing half the projected number of stars, estimated in \cite{Vitral21}.} Furthermore, the median and maximum \hst\ proper motion baselines are of 10.6 and 12.6 years for this cluster, yielding even better data than for NGC~3201 and NGC~6397, which we analysed previously.

The Jeans modelling performed by \mpo, as in many other routines (see \citealt{Read+21} for a comparison of different mass-modelling algorithms), does not take into account the rotation of the system. In the case the system does have significant rotation, the tangential anisotropy might be overestimated, as well as the cluster's internal velocity dispersion, which could translate to poor mass fits. Fortunately, the M4 internal rotation profile is relatively low, with several analyses assigning negligible plane-of-sky rotation \citep[i.e., values $\lesssim 5\%$ of the velocity dispersion][]{Bianchini+18,Vasiliev19c,Sollima+19}. There is more rotation in the line-of-sight data \citep[$\sim 10\%$ of the velocity dispersion, according to][]{Malavolta+15}, but this still remains low when compared to the recently analysed NGC~3201 in Paper~I and 47~Tuc in \cite{Mann+19}, where the maximum ratio of rotation velocity by velocity dispersion in the plane-of-sky reaches $0.10$ and $0.51$ respectively, according to Table~1 from \cite{Bianchini+18}. In fact, the velocity dispersion and amplitude of the rotation profile of M4 resembles that of NGC~6397, where it has been previously argued that the rotation was not enough to significantly affect mass measurements \citep[e.g.,][]{Kamann+16}.  Another caveat is that \mpo\ considers the system spherical, which once again suits the case of M4 very well, since its ratio of semi-minor ($b$) by semi-major ($a$) axis of the projected ellipse are of the order of $0.95$ (\citealt{White&Shawl87} and  \citealt{Chen&Chen10} estimate $b/a = 1.00 \pm 0.01$ and $0.93\pm0.02$, respectively).

The inner stellar density profile of M4 (i.e., $\sim 0.1$~pc) is important for understanding the contents of its core.  Since it does not present the characteristic core-collapsed inner cusp at very inner radii 
(such as for NGC~6397, for $R/r_{\rm h}<0.1$ in Figure~\ref{fig: cc-comp}), M4 has  so far been considered as a non core-collapse cluster by most studies targeting this structural shape in GCs \citep[e.g.,][]{Djorgovski&King86,Trager+95,McLaughlin&vanderMarel05}.
However, M4 remains a very dense cluster (see Figure~\ref{fig: cc-comp} for a comparison) 
that might be close to reaching 
core-collapse. Indeed, \cite{Heggie&Giersz&Giersz08} used Monte Carlo simulations to propose that M4 had already reached a post core-collapse phase, and its core was being sustained by \textit{binary burning}, although the clear lack of the inner cusp mentioned above still renders this result debatable 
(Section~\ref{ssec: tangential} also argues against core-collapse in this cluster). 
If M4 is close to reaching core-collapse, then one could expect that it would have a small, yet not negligible black hole population (see introduction of Paper~I as well as \citealt{Morscher2015,wang2016dragon,Askar2017,Kremer+20,Rodriguez+22}).\footnote{If M4 did not form black holes, then the cluster evolution could be different than discussed. However, the zero black holes possibility would require a \textit{highly} non-standard initial mass function \citep[e.g.,][]{Weatherford+21}. Although this possibility cannot be completely ruled out, consideration of this is beyond the scope of this paper.}  This BH population would behave as a very dense sub-clustered dark mass in the cluster's core.  At the same time, other structural parameters of M4 are, in general, very similar to its core-collapsed counterpart NGC~6397, being a relatively small cluster (3D half-mass radius 3.95~pc and a mass of $9.04\times10^4$~M$_{\odot}$, according to the website of H.~Baumgardt\footnote{\url{https://people.smp.uq.edu.au/HolgerBaumgardt/globular/}, \copyright\ H.~Baumgardt, A.~Sollima, M.~Hilker, A.~Bellini \& E.~Vasiliev \citep{Baumgardt17,Baumgardt+19,Baumgardt+20,Baumgardt&Vasiliev21,Sollima&Baumgardt17,Vasiliev&Baumgardt&Baumgardt21}. \label{fn: Baumgardt}}) and very dense.

\begin{figure}
\centering
\includegraphics[width=0.95\hsize]{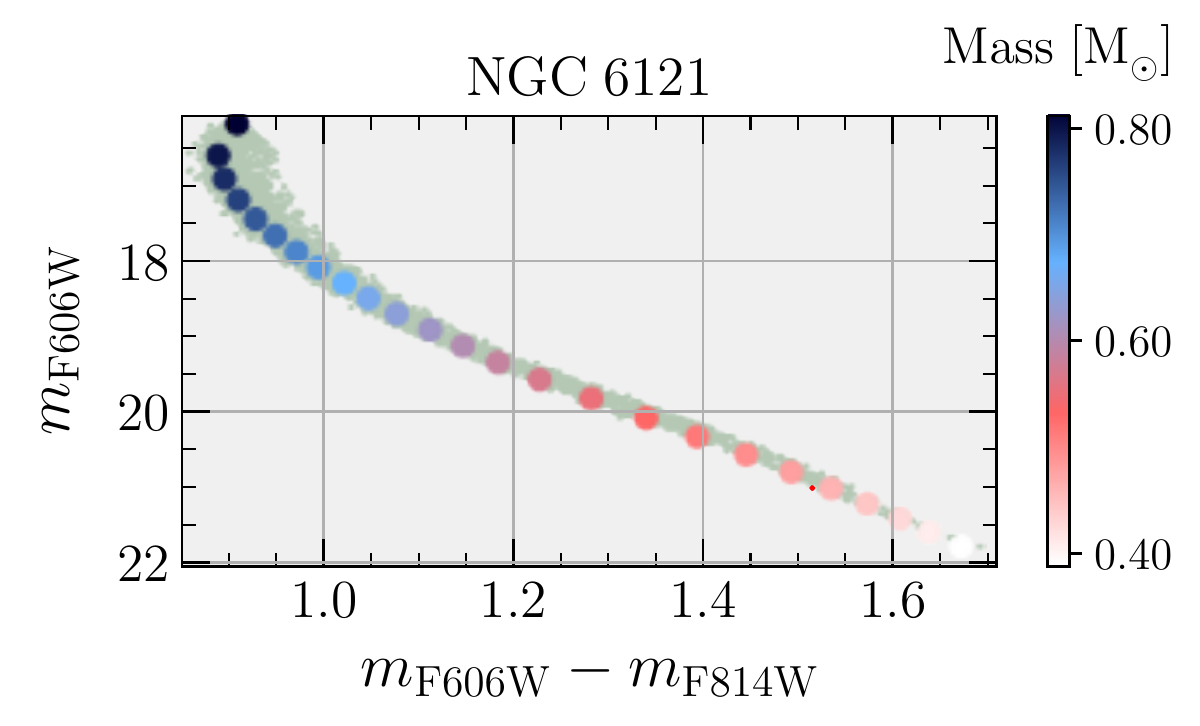}
\caption{\textit{Colour-magnitude diagram:}
The \emph{small gray-green points} are the \hst\ data, cleaned according to Sect.~\ref{ssec: cleaning}, while the \emph{filled circles} are the predictions from the \parsec\ code, colour-coded by stellar mass.}
\label{fig: parsec-iso}
\end{figure}

%
\subsection{Data cleaning} \label{ssec: cleaning}
%
%
The \hst\ and \gaia\ EDR3 data cleaning in this paper follows the procedures explained in detail in Paper~I (Section~3), with previous checks performed also in \cite{Libralato+19}.  The cleaning consisted mostly in defining well-measured thresholds for photometric and astrometric flags, along with a proper motion error threshold being smaller than the local \footnote{We used a network of the hundred closest stars in projected radius and mass (see Paper~I).} velocity dispersion and further filtering field stars in proper motion and colour magnitude spaces.  Given this, we focus here on sharing the values of specific physical quantities used for M4 and on describing the information about this cluster that will be useful in our analysis. In particular, concerning the equivalent of Table~1 of Paper~I, we list below the respective information used for M4:

\begin{itemize}
    \item Distance to the Sun: 1.85 kpc \citep{Baumgardt&Vasiliev21}, yielding $1''=9.0\,\rm mpc$;
    \item Reimers scaling factor (i.e., Red Giant Branch mass-loss efficiency): 0.402 \citep{McDonald&Zijlstra15};
    \item Age: 12.74 Gyr \citep{MarinFranch+09};
    \item Total extinction, considering $R_{\rm v} = 3.1$: 1.3262 \citep{Schlafly&Finkbeiner11}\footnote{Notice that M4 has a non-standard extinction coefficient of $R_V = 3.76$ \citep{Hendricks+12}. Nonetheless, this should not affect our results since this information is required only when converting magnitudes to mass through \parsec\ isochrones, in order to obtain a close net of stars for the proper motion error cleaning (see Paper~I, section~3.3). Because we tested different error thresholds in Section~\ref{ssec: robustness} (see lines 13--16 from Table~\ref{tab: results-full}), we show that our analyses are not significantly affected by this point.};
    \item Metallicity, in log solar units: --0.83 \citep{MarinFranch+09};
    \item Cluster centre, in degrees: ($245\fdg89669$, $-26\fdg52584$) (\citealt{Vitral21}, calculated with \balrogo\footnote{\balrogo\ estimates the cluster centre by considering the centre of mass of stellar counts, fitted in a Bayesian framework with a Plummer \citep{Plummer1911} density profile.}).
\end{itemize}

With part of the information above, we constructed the \parsec\ isochrone \footnote{\url{http://stev.oapd.inaf.it/cgi-bin/cmd}} (e.g., \citealt{Bressan+12,Chen+14,Chen+15,Marigo+17,Pastorelli+19}) displayed in Figure~\ref{fig: parsec-iso}. This indicates that the visible Main Sequence stars have masses below $0.80\,\msun$.

The only cleaning procedure that was performed differently from Paper~I concerned the colour-magnitude diagram (CMD) interloper filtering, for the specific case where the maximum proper motion error threshold was set to be half of the local velocity dispersion. In this case, we noticed that setting a 2-$\sigma$ confidence contour for the Kernel Density Estimation (KDE) as in the Paper I was too conservative:  it removed stars that clearly belonged to the cluster's CMD. This happens because as the dataset gets smaller (as a direct result from the more conservative error threshold), the original KDE contour tends to prune the faintest parts of the main-sequence more severely than necessary. In order to keep such stars while still filtering the dataset against interlopers, non-resolved binaries and blue stragglers, it sufficed to slightly increase the 2-$\sigma$ contour up to a 2.5-$\sigma$ limit. As the original subset (with the standard error threshold used in Paper~I) 
was not severely pruned by the KDE contour on the fainter magnitude end, this problem was not observed and we decided to keep the analysis as in Paper~I, for consistency.

\subsection{Mass modelling} \label{ssec: mass-mod}

%
The mass modelling tools used throughout the paper follow Section~4 from Paper~I. We briefly 
reiterate the most salient points below, and direct the reader to the 
Paper I for further details.

\subsubsection{Jeans modelling} \label{sssec: jeans}

We employ the Bayesian mass-orbit modelling code \mpo\ \citep[][Mamon \& Vitral in prep.]{Mamon+13,Read+21} to estimate masses, the velocity anisotropy and parameters from the cluster density profile. The velocity anisotropy (`anisotropy' for short) is defined as in \citet{Binney80}:
\begin{equation}    \label{eq: anisotropy}
    \beta(r) = 1 - \displaystyle{\frac{\sigma_{\theta}^{2}(r) + \sigma_{\phi}^{2}(r)}{2 \,\sigma_{r}^{2}(r)}} \ ,
\end{equation}
where $\theta$ and $\phi$ are the tangential components of the coordinate system, while $\sigma_{i}^2$ stands for the velocity dispersion of the component $i$ of the coordinate system. In spherical symmetry, $\sigma_\phi = \sigma_\theta$.
To adjust the data, \mpo\ assumes that the \emph{local} velocity ellipsoid is an anisotropic Gaussian, whose major axis is aligned with the spherical coordinates.

Then, \mpo\ fits parametric models for the radial profiles of total mass and the velocity anisotropy of the visible stars to the distribution of these stars in projected phase space. It does so by solving the spherical, stationary, Jeans equation 
with no streaming motions
\citep{Binney80}    
\begin{equation} 
  \frac{{\rm d}\left (\rho\sigma_r^2\right)}{{\rm d}r} + 2\,\frac{\beta(r)}{
    r}\,\rho(r)\sigma_r^2(r) = -\rho(r) \frac{G\,M(r)}{r^2} \ ,
\label{eq: jeans}
\end{equation}
for the radial velocity dispersion profile, $\sigma_r(r)$,
assuming a given mass profile $M(r)$ and anisotropy profile $\beta(r)$, for a previously determined mass density profile 
$\rho(r)$ for the kinematic tracers (here stars). The term $\rho\,\sigma_r^2$ is the dynamical pressure that counteracts gravity.\footnote{In fact, the Jeans equation~(\ref{eq: jeans}) is a consequence of the Collisionless Boltzmann Equation, which considers the incompressibility in phase space of the six-dimensional (6D) distribution function (DF). Expressing the distribution function in terms of a 6D number, mass or luminosity density, implies that the term $\rho$ in the Jeans equation is the number, mass or luminosity density. For the present case of a globular cluster made of stars, it makes more physical sense to work with mass density. In the absence of mass segregation, the mass density is proportional to the number density, so the mass density profile is obtained from deprojecting the observed surface number density profile.} The anisotropic runs of \mpo\ used the generalisation (hereafter gOM) of the Osipkov-Merritt model \citep{Osipkov79, Merritt85} for the velocity anisotropy profile:
\begin{equation}    \label{eq: gOM}
    \beta_{\mathrm{gOM}}(r) = \beta_{0} + (\beta_{\infty} - \beta_{0}) \ \displaystyle{\frac{r^2}{r^2 + r_{\beta}^2}} \ ,
\end{equation}
where $r_{\beta}$ is the anisotropy radius, which can be fixed as the scale radius of the luminous tracer by \mpo.\footnote{\cite{Mamon+19} and \cite{Vitral&Mamon21} found no significant change in models of galaxy clusters and globular clusters, respectively, when using this model for $\beta(r)$ compared to one with a softer transition: $\beta(r) = \beta_{0} + (\beta_{\infty} - \beta_{0})\,r/(r+r_{\beta})$, first used by \cite{Tiret+07}.} 

The number density profile, $\nu(r)$ (assumed proportional to the stellar mass density profile $\rho(r)$), is determined from the surface density profile, assuming spherical symmetry. We fit the distribution of projected distances $R$ to the photometric data (since the proper motion data is incomplete in the inner regions) using
the S\'ersic profile \citep{Sersic63,Sersic68} as in \cite{Vitral&Mamon21} and Paper~I, which simultaneously fits the 
cluster's extent and density inner slope, allowing for a higher degree of freedom. The deprojection of this profile into 3D coordinates uses the same method as described in appendix~A from \citeauthor{Vitral&Mamon21} (\citeyear{Vitral&Mamon21}, which employs the analytical forms from \citealt*{LimaNeto+99},  \citealt{Simonneau&Prada04},  and \citealt{Vitral&Mamon20}).
We iterate once using \mpo\ with priors from our first step (see Paper~I), where we test if different mass models will not yield very distinct priors. If it is the case, we thus select the most liberal priors among all.

Given that the proper motion data is not complete in projected distance, the likelihood is written in terms of probabilities of plane-of-sky velocities at given projected distance:
\begin{equation}
    {\cal L} = \prod_i p({\bf v}_\mathbf{i}\,|R_i) \ .
\end{equation}
Thus, the conditional probability of measuring a velocity ${\bf v}_\mathbf{i}$ is the mean of the local  velocity distribution function, $h({\bf v}\,|\,R,r)$, integrated along the line of sight:
\begin{equation}
    p({\bf v}\,|\,R) = \frac{2}{\Sigma(R)}\, \int_R^\infty h({\bf v}\,|\,R,r)\,\nu(r) \,\frac{r}{\sqrt{r^2-R^2}}\,{\rm d}r \ .
    \label{pvofR}
\end{equation}
\mpo\ assumes that the local velocity distribution functions are Gaussians in spherical coordinates. \mpo\ determines the marginal distributions of the free parameters and their covariances by running the Markov Chain Monte Carlo (MCMC) routine ({\sc CosmoMC},\footnote{\url{https://cosmologist.info/cosmomc/}.} \citealt{Lewis&Bridle02}).  
We ran 6 MCMC chains in parallel, which move around the multi-dimensional parameter space following the Metropolis-Hastings algorithm, which produces distributions of parameter values proportional to their posteriors. Thus the MCMC chain elements (past an initial burn-in phase that remembers the initial choice of parameters) produce a statistical description of the posteriors of each parameter and their correlations.
We generally use flat priors on log mass, log scale radii, and on the symmetrised anisotropy parameter $\beta_{\rm sym} = \beta/(1-\beta/2)$, and Gaussian priors on the pre-determined surface density profile parameters (S\'ersic index and log effective radius) and on the bulk motions.

\subsubsection{Astrometric handling}

We use the astrometric routines from \balrogo\footnote{\url{https://gitlab.com/eduardo-vitral/balrogo}} \citep{Vitral21} in order to derive fits and constraints on the cluster surface density, centre and bulk proper motion, and to assign membership probabilities between stars from the cluster and from the field. In particular, \balrogo\ assigns a fat-tailed, asymmetric Pearson~VII \citep{Pearson16} distribution to the proper motion distribution of Milky Way contaminants,\footnote{This Pearson~VII distribution was originally proposed by \cite{Vitral&Mamon21}, while \balrogo\ generalised it to being asymmetric.} and a Gaussian to the cluster members, which allows us to compute membership probabilities to each star.

\subsubsection{Statistical tools}

The statistical methods used to select between different mass models also follow the description of Paper~I, including the construction of GC mocks with the \agama\ software \citep{Vasiliev19a} in the same fashion as previously (see their Section~4.2). The comparison of the \mpo\ fits of mock and true datasets uses (again as in Paper~I) 
the fraction of MCMC chain elements whose absolute difference is greater than that for the best likelihood solutions of mock and observed fits (higher fractions indicate better agreement between mock and observed preferred solutions). Similarly, we also use Kolmogorov-Smirnov (\citealt{Kolmogorov1933,Smirnov1939}, hereafter KS) as well as Anderson-Darling (\citealt{Anderson&Darling52}, hereafter AD)  statistics to quantify the disagreement between mock and observed marginal distributions of mass and scale radius of a dark central component (smaller KS and AD statistics indicate better agreement between the different marginal distributions). 

Finally, we use Bayesian evidence methods that penalise the likelihood for extra free parameters. 
In particular, we adopted the corrected Akaike Information Criterion (derived by \citealt{Sugiyara78} 
and independently by \citealt{HurvichTsai89} who demonstrated its utility for a wide range of models)
\begin{equation}
      \mathrm{AICc} =\mathrm{AIC} + 2 \, \frac{N_{\mathrm{free}} \,  (1 + N_{\mathrm{free}})}{N_{\mathrm{data}} - N_{\mathrm{free}} - 1} \ ,
\end{equation}
where AIC is the original 
{Akaike Information Criterion}
\citep{akaike1973information}
\begin{equation}
 \label{eq: AIC}
    \mathrm{AIC} = - 2 \, \ln \mathcal{L_{\mathrm{MLE}}} + 2 \, N_{\mathrm{free}} \ ,
\end{equation}
and where ${\cal L}_{\rm MLE}$ is the maximum likelihood estimate found when exploring the parameter space, $N_{\rm free}$ is the number of free parameters, and $N_{\rm data}$ the number of data points. The likelihood (given the data) of one model relative to a reference one is 
\begin{equation}
\rm \exp\left(-\frac{AIC-AIC_{\rm ref}}{2}\right)
\label{eq: pAIC}
\end{equation}
\citep{Akaike83,Burnham&Anderson02} and we assume strong evidence for one reference model over another whenever $95$  per cent confidence is attained (i.e., $\rm AICc > AICc_{\rm ref}+6$). We consider AICc differences smaller than $4.5$ (i.e., less than $90$  per cent confidence) are usually not enough to consistently distinguish two models, based on purely statistical arguments (thus, no astrophysics involved). Specific details are provided in Section~4.3 of Paper~I.

\subsection{Monte Carlo models} \label{sssec: monte-carlo}

To assist our interpretation of the results from the Jeans modelling analyses, we followed Paper~I, using Monte Carlo evolutionary $N$-body models constructed with the cluster dynamics code \cmc\ \citep{Kremer+20,Rodriguez+22}.  \texttt{CMC} is a H\'{e}non-type Monte Carlo code that includes various physical processes relevant to the dynamical evolution of GCs, including two-body relaxation, tidal mass loss, and direct integration of small-$N$ resonant encounters.
By employing the \texttt{COSMIC} single/binary star evolution code \citep{Breivik+20}, \texttt{CMC} tracks various evolution features (including stellar type, mass, radius, luminosity, etc.) for all $N$ stars as the model cluster evolves dynamically. 
This makes it straightforward to compute standard observed cluster features from the \texttt{CMC} snapshots, in particular surface brightness and velocity dispersion profiles, binary fractions, and colour-magnitude diagrams.
See \cite{Rui+21b} for a more detailed explanation of how the models are matched to existing clusters.

As M4 has not yet been analysed in previous \cmc-related analyses, we computed new models, which were later matched to the M4 surface brightness and velocity dispersion profiles from \cite{Trager+95}\footnote{Figure~\ref{fig: sbp-cmc} depicts this match.} and \cite{Baumgardt&Hilker18}, respectively. 
In particular, our preferred model started with a \emph{virial radius}\footnote{The `virial radius' of a GC is defined here as in \cite*{Binney2008,PortegiesZwart2010} and \cite{Kremer+20p}, i.e. $r_{\rm v} = GM^2 / (2\,|U|)$, where $U$ is the total cluster potential energy, $M$ its mass and $G$ is the gravitational constant.  
\label{ftn: rv}} of 1~pc, initial metalliticy $Z = 0.002$, initial binary fraction of $5\%$ and initial Galactocentric position of 8~kpc, which are values similar to the ones known for M4 \citep[e.g.,][]{Baumgardt&Hilker18,MarinFranch+09,Milone+12_binaries,GaiaHelmi+18}.
Indeed, more evolved clusters, with smaller initial virial radii, tend to reach the core-collapse phase sooner by ejecting most of its original black hole population \citep[e.g.,][]{Kremer+20}, while clusters with significantly greater initial virial radius show a much larger 
core than the one observed in M4's density profile (and also retain many black holes). 


\section{Results \& robustness} \label{sec: results}

\begin{figure}
\centering
\includegraphics[width=0.98\hsize]{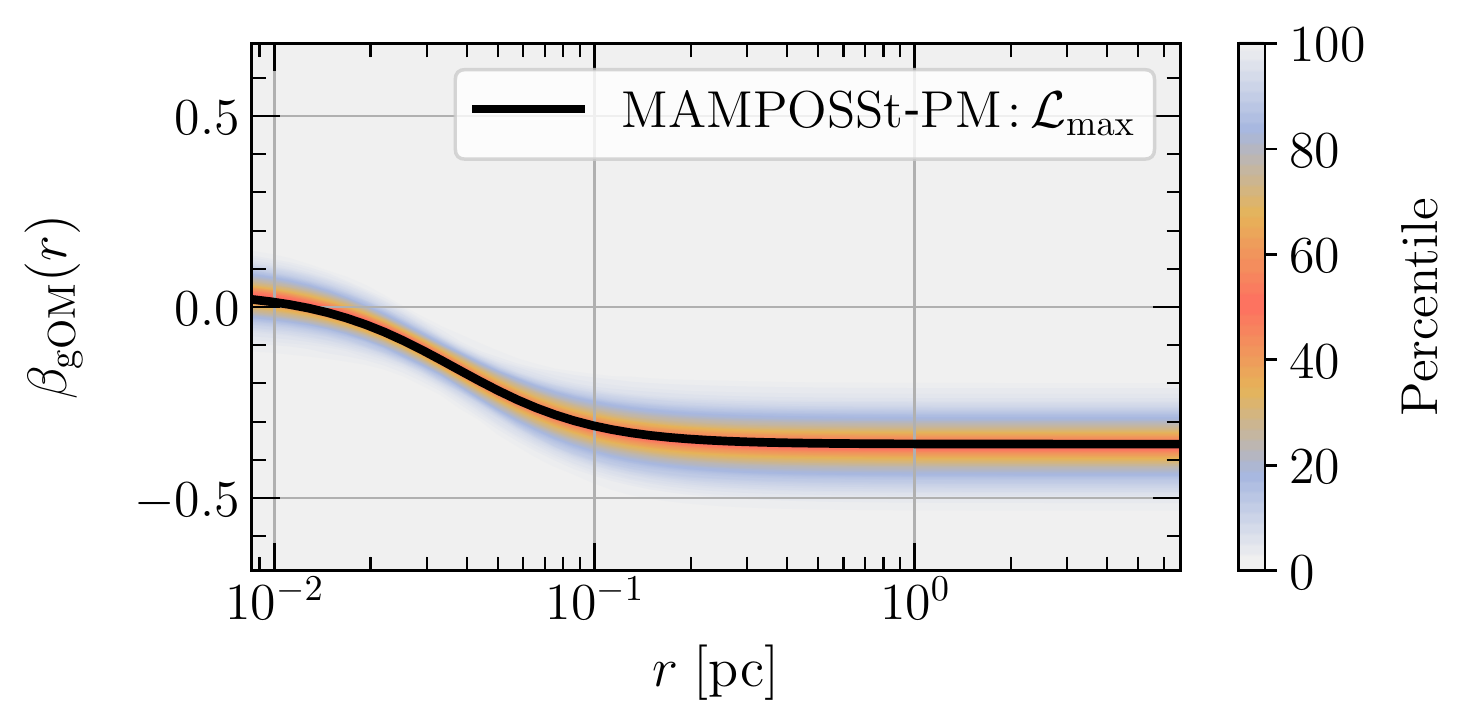}
\caption{\textit{Velocity anisotropy:} \mpo\ fits of the velocity anisotropy, using the Osipkov-Merritt parameterization \protect\citep{Osipkov79,Merritt85}, generalised to free inner and outer anisotropies (gOM), as a function of the 
physical distance to the cluster centre. The colour bar indicates the percentile of the MCMC chain post burn-in phase. The \emph{black curve} represents the maximum likelihood solution of our fit.
The range of physical radii is set to the range of projected radii in the data we analysed.}
\label{fig: anis-mpo}
\end{figure}

\subsection{Velocity anisotropy} \label{ssec: anis}

The mass-modelling routine in \mpo\ allows the user to fit the velocity anisotropy profile (Eq.~[\ref{eq: anisotropy}] 
with the gOM parameterisation of Eq.~[\ref{eq: gOM}]) of the studied tracers as a function of the distance from the cluster's centre, $r$. This is because the anisotropy appears during the solving of the Jeans equation,\footnote{The use of two-dimensional proper motions, such as in our dataset, allows us to break the known mass-anisotropy degeneracy \citep[e.g.][]{Binney&Mamon82}.} as depicted in Section~\ref{sssec: jeans}.
Differently from our previous analysis of NGC~3201 and NGC~6397, where isotropy was preferred\footnote{Moderate radial anisotropy was measured at outermost radii, but with no statistical significance.} through the whole extent of our data, the case of M4 shows a robust signal of tangential anisotropy at intermediate and large radii: $\beta = -0.4\pm0.1$, which amounts to $\sigma_\theta/\sigma_r = 1.18\pm0.04$. Yet, the very inner regions remain strongly isotropic (see Figure~\ref{fig: anis-mpo}). This result qualitatively agrees with the projected anisotropy profile measured by \cite{Vasiliev&Baumgardt&Baumgardt21}.
Their milder velocity anisotropy is the consequence of their measuring the \emph{projected} anisotropy compared to the \emph{three-dimensional} anisotropy determined by \mpo.
%
Indeed, the random mixing of different line-of-sight layers renders the projected velocity anisotropy more isotropic, or in this particular case, less tangential.

The statistical significance of this result is also remarkable, with the models allowing for a free anisotropy fit displaying AICc values much smaller than the fits with fixed isotropy (i.e., $\Delta \rm AICc \sim 30$). In summary, the probability of full isotropy over outer tangential anisotropy (with inner isotropy) is less than $10^{-6}$. Hence, in all of our further \mpo\ fits, we do not force an isotropic profile, thus better modelling the tangential anisotropy and its possible impacts on our mass estimates.

\begin{figure*}
\centering
\includegraphics[width=0.247\hsize]{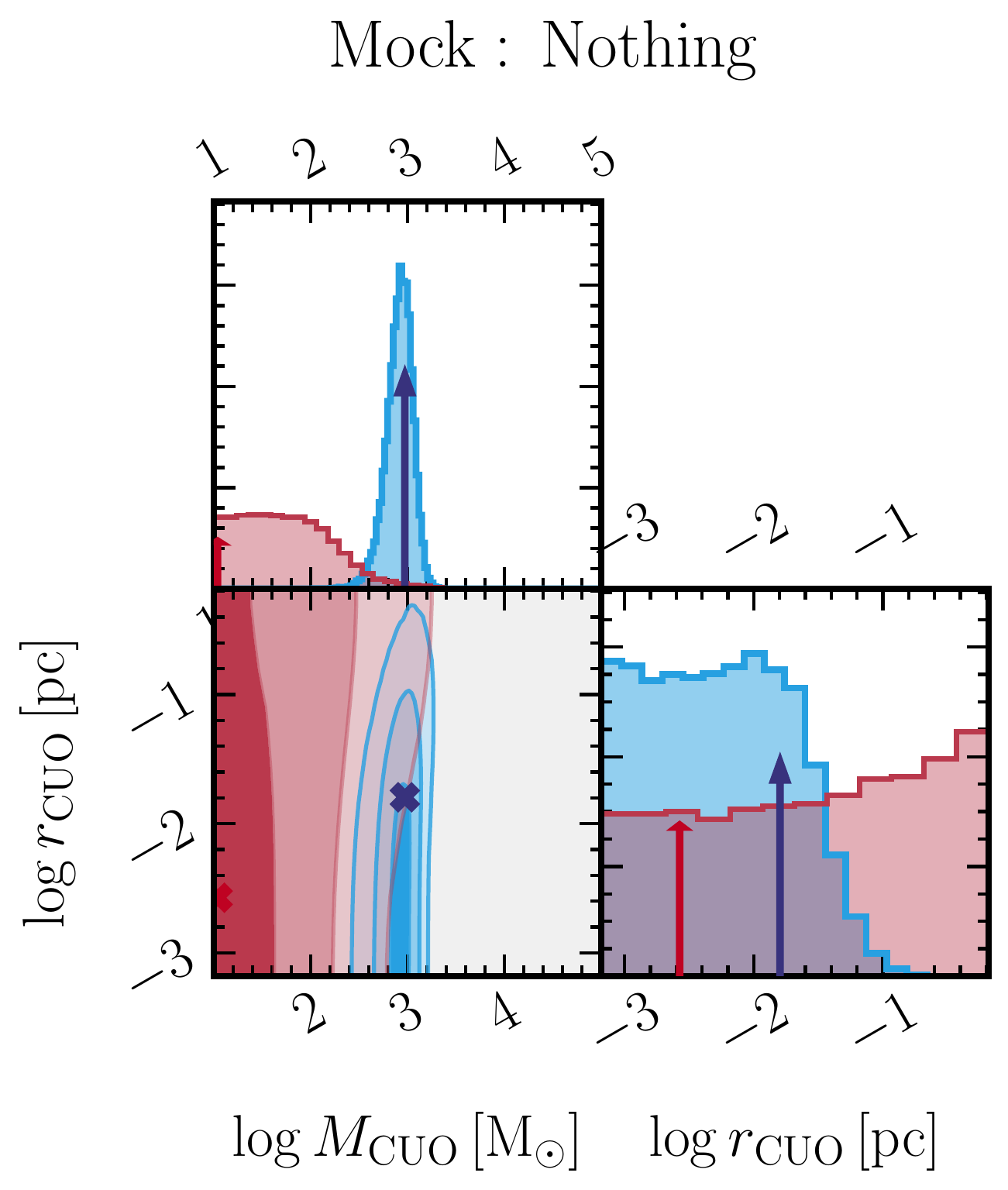}
\includegraphics[width=0.247\hsize]{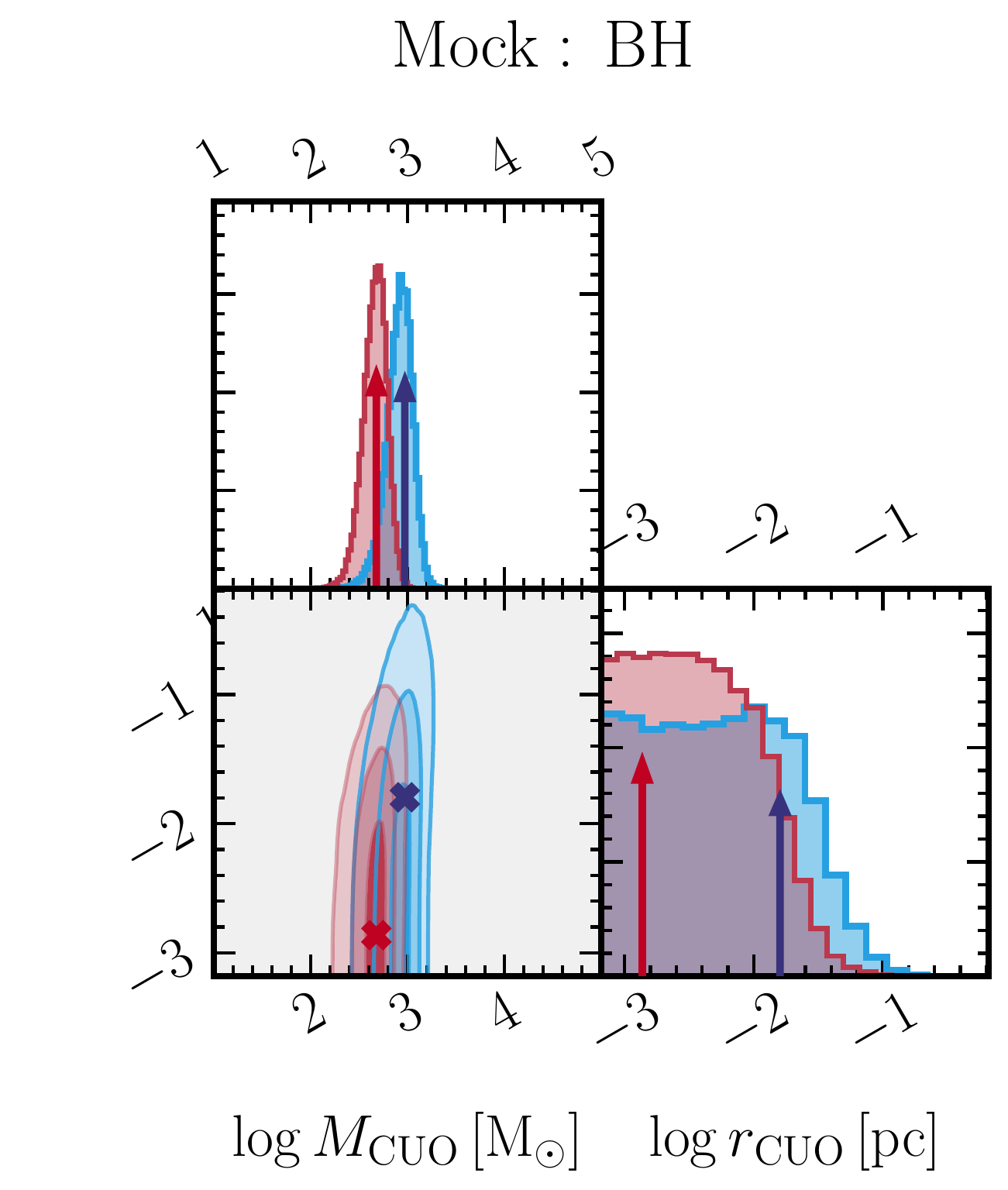}
\includegraphics[width=0.247\hsize]{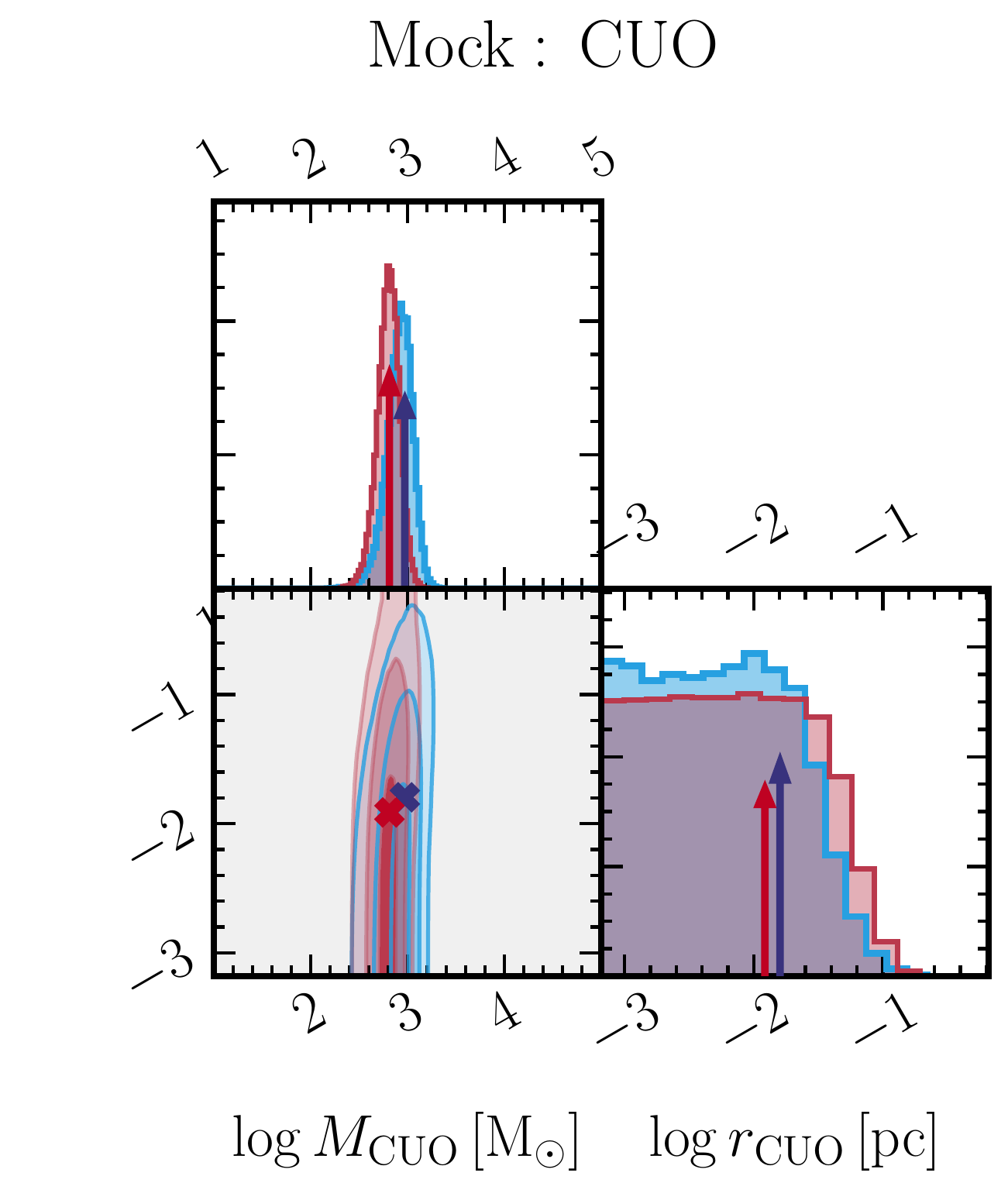}
\includegraphics[width=0.247\hsize]{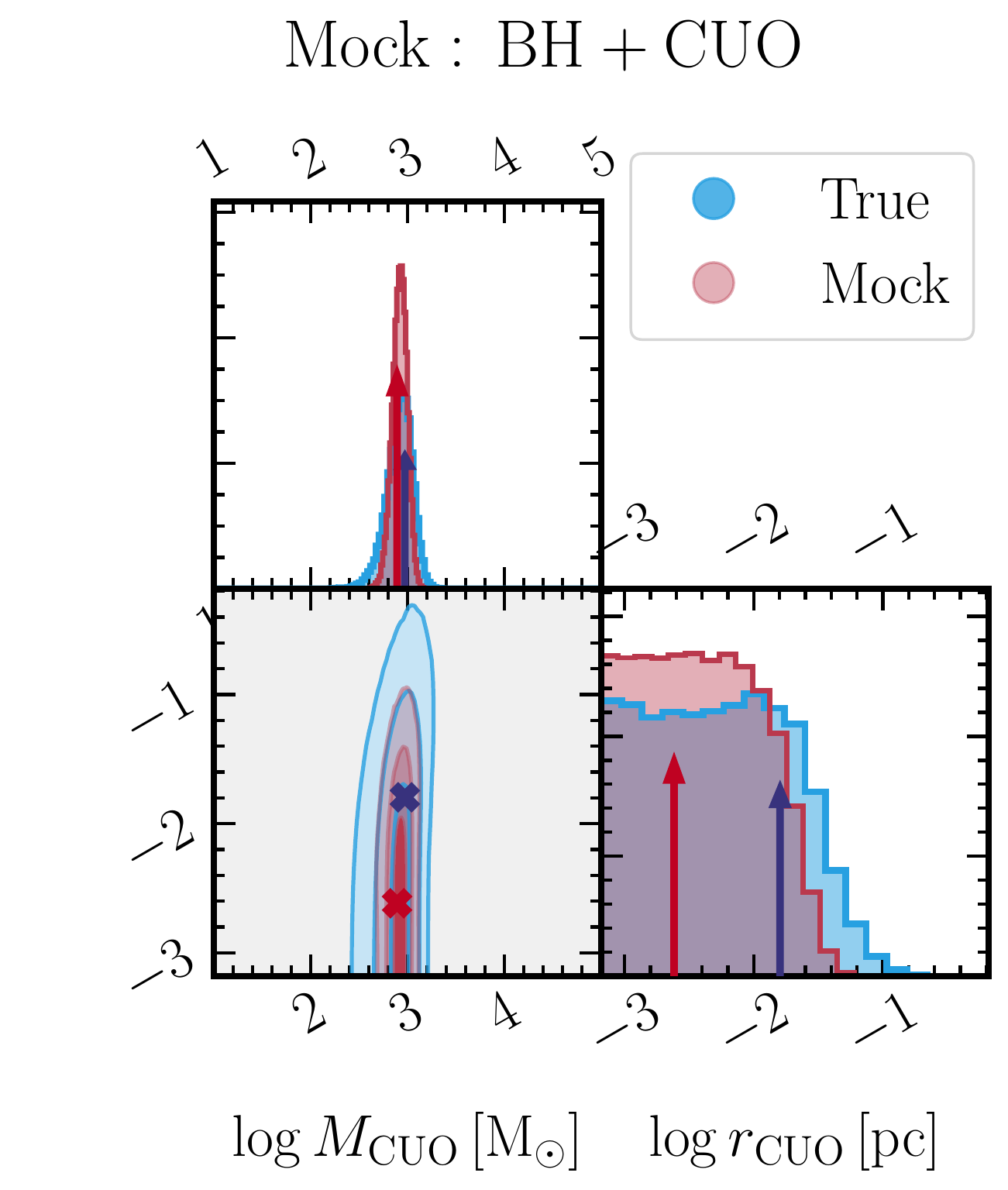}
\caption{{\it Mock data comparison:} Marginal distributions of the cluster of unresolved objects (CUO) mass and 2D Plummer half mass radius and their covariances for the true data (\hst\ and \gaia\ EDR3) in \textit{blue} and the mock data (constructed with \agama) in \textit{red}. 
The priors are flat for $\log M_{\rm CUO}$ within the plotted range and zero outside, while they are Gaussian for the $\log$ scale radii, centred on the middles of the panels and extending to $\pm3\,\sigma$ at the edges of the panels, and zero beyond. The arrows indicate the respective best likelihood solutions of the MCMC chains.
The mock data prescription is, from \textbf{left} to \textbf{right}: No central dark component (Nothing); a central black hole alone (BH); a central CUO (CUO) and both a central black hole and CUO (BH$+$CUO). The mocks were constructed with the best values of each respective mass model (lines 5--8) from Table~\ref{tab: results-full}. The fits alone indicate a preference for a central dark mass in M4.
}
\label{fig: comp-standard}
\end{figure*}

\begin{figure*}
\centering
\includegraphics[width=0.99\hsize]{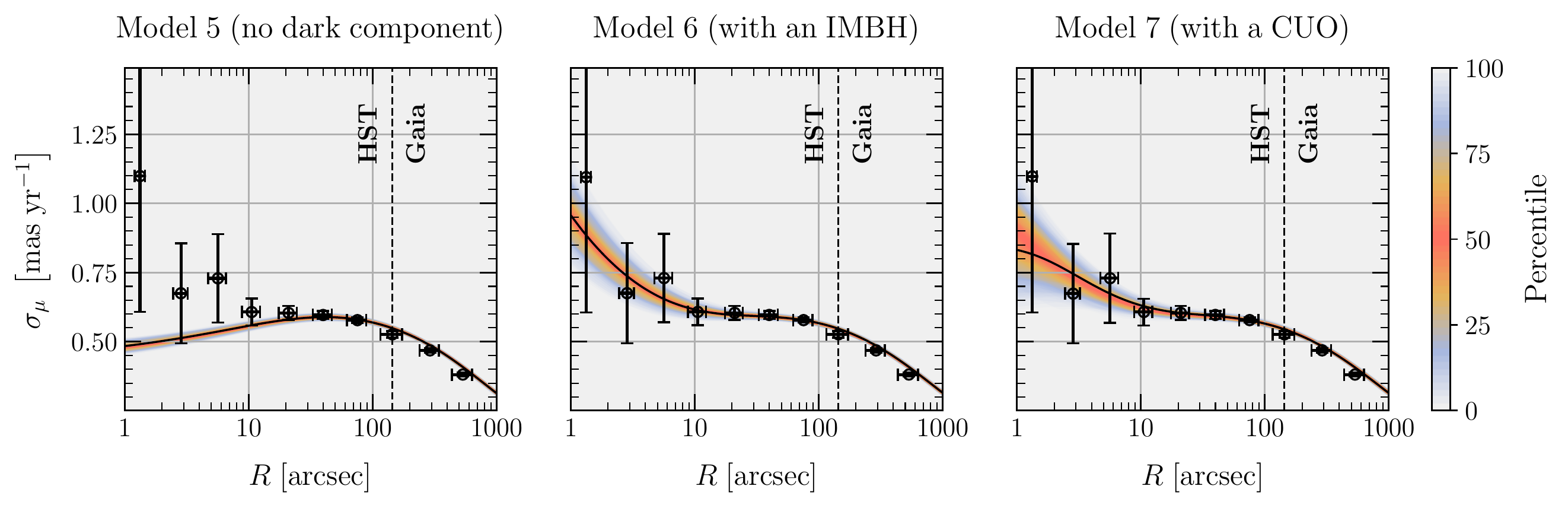}
\caption{{\it Goodness of fit:} Display of the proper motion velocity dispersion as a function of the projected radius
for models~5 (no dark component, {\bf left}), 6 (IMBH, {\bf middle}) and
7 (central unresolved objects, {\bf right}). 
The \textit{coloured regions} represent the percentiles from the MCMC chains of our fit, while the continuous \textit{black solid line} shows the maximum likelihood solution from \mpo. The \textit{black dashed line} depicts the region separating the \hst\ and \gaia\ stars used in our modelling.
The \emph{black circles} and vertical error bars feature the measured proper motion dispersion and respective 1-$\sigma$ uncertainty, calculated with the recipe from \protect\citeauthor{vanderMarel&Anderson&Anderson10}~(\citeyear{vanderMarel&Anderson&Anderson10}, appendix~A), in ten logarithmically-spaced radial bins. The horizontal error bars considered the 1-$\sigma$ radial quantization noise. The plot highlights the need for a very concentrated central dark mass in M4.
}
\label{fig: vdisp}
\end{figure*}

\subsection{Dark central mass} \label{ssec: dark-mass}
%

\begin{table}
\caption{Main statistical tests used for model selection}
\label{tab: statistics}
\centering
\renewcommand{\arraystretch}{1.2}
\tabcolsep=2.5pt
\footnotesize
\begin{tabular}{l@{\hspace{3mm}}lccrrrr}
\hline\hline   
\multicolumn{1}{c}{Test ID} &
\multicolumn{1}{c}{Mock} &
\multicolumn{1}{c}{$\phi$} &
\multicolumn{1}{c}{$\phi$} &
\multicolumn{1}{c}{AD} &
\multicolumn{1}{c}{AD} &
\multicolumn{1}{c}{KS} &
\multicolumn{1}{c}{KS} \\
\multicolumn{1}{c}{} &
\multicolumn{1}{c}{model} &
\multicolumn{1}{c}{$M_{\rm dark}$} &
\multicolumn{1}{c}{$r_{\rm dark}$} &
\multicolumn{1}{c}{$M_{\rm dark}$} &
\multicolumn{1}{c}{$r_{\rm dark}$} &
\multicolumn{1}{c}{$M_{\rm dark}$} &
\multicolumn{1}{c}{$r_{\rm dark}$} \\
\multicolumn{1}{c}{(1)} &
\multicolumn{1}{c}{(2)} &
\multicolumn{1}{c}{(3)} &
\multicolumn{1}{c}{(4)} &
\multicolumn{1}{c}{(5)} &
\multicolumn{1}{c}{(6)} &
\multicolumn{1}{c}{(7)} &
\multicolumn{1}{c}{(8)} \\ 
\hline
$\beta(r)$ & Nothing    & 3\%  & 58\% & 57592 & 3633  & 0.373 & 0.085 \\
$\beta(r)$ & IMBH       & 41\% & 15\% & 270   & 91    & 0.018 & 0.018 \\
$\beta(r)$ & CUO        & 47\% & 89\% & 1070  & 108   & 0.042 & 0.016 \\
$\beta(r)$ & IMBH$+$CUO & 59\% & 28\% & 11871 & 148   & 0.122 & 0.015 \\
\hline
$v_{\star}$ & Nothing    & 26\% & 45\% & 21765 & 4412  & 0.190 & 0.111 \\
$v_{\star}$ & IMBH       & 98\% & 84\% & 8809  & 1166  & 0.109 & 0.056 \\
$v_{\star}$ & CUO        & 77\% & 93\% & 810   & 1542  & 0.061 & 0.034 \\
$v_{\star}$ & IMBH$+$CUO & 93\% & 67\% & 846   & 20757 & 0.043 & 0.221 \\
\hline
\end{tabular}
\parbox{\hsize}{\textit{Notes}: 
The statistical tests compare \mpo\ outputs (with CUO prior) on mocks and on data.
Columns are 
\textbf{(1)} test ID (according to column~3 from Table~\ref{tab: results-full}); 
\textbf{(2)} mass model assigned to the mock data; 
\textbf{(3)} fraction of MCMC chain elements that present absolute differences in dark mass  greater than that between the mock and true data fit's most likely solutions -- higher values indicate good agreement between the mock and true data fits;
\textbf{(4)} same as (3), but considering  the dark radius;
\textbf{(5)} AD statistic, for $M_{\rm dark}$ -- high values indicate poor matches; 
\textbf{(6)} AD statistic, for $r_{\rm dark}$; 
\textbf{(7)} KS statistic, for $M_{\rm dark}$ -- high values indicate poor matches; 
\textbf{(8)} KS statistic, for $r_{\rm dark}$.
}
\end{table}

As in \cite{Vitral&Mamon21} and Paper~I, we test four different mass configurations with \mpo: 1) no central dark component (Nothing); 2) a central, single black hole (BH); 3) an inner \emph{cluster of unseen objects} (CUO), and 4) central black hole plus a CUO (BH$+$CUO). Our standard anisotropic runs, displayed in lines 5--8 of Table~\ref{tab: results-full} point very clearly to the presence of a central mass of roughly 800~M$_{\odot}$. The differences in AICc indicate that the model with no central dark component has a probability of less than $0.005\%$ percent when compared to the model with both CUO and BH, of $0.002\%$ when compared to the CUO model, and only of $0.0008\%$ when compared to the model with a single central black hole, of mass $792^{+253}_{-217}\,\msun$ (best AICc value). 
The distinction of AICc values among the models with a central dark mass is however, too small to yield significant statistics. We therefore follow Paper~I and use mock datasets constructed with \agama\ to probe the similarities of outputs from different mass models fitted by \mpo.

Figure~\ref{fig: comp-standard} compares the marginal distributions of CUO mass and scale radius and their covariance for the \mpo\ fit to mock data (light red) to the \mpo\ fit to the observed data (light blue), both assuming the CUO mass model. 
As argued in Paper~I, whether or not the CUO model is the correct mass model, one expects that the marginal distributions of the CUO scale radii should have similar shapes when comparing those obtained on a mock that represents the observed data and those directly obtained from the same data.
The figure also compares the values of the maximum likelihood estimates (arrows)
of the CUO log mass and scale radius for both mock and observed data.
. 

First, it is evident that a mock simulation with no dark central mass (`Nothing') has a completely different mass marginal distribution than the one observed in the true data, once again attesting to the robustness of a central mass excess in M4.  
However, when trying to select the best mass excess model from the agreement of $\log{r_{\rm CUO}}$ marginal distributions, the choice is less evident. Similarly to NGC~3201 in Paper~I, the overall marginal distribution shapes are very similar among models BH, CUO and BH$+$CUO, while the agreement in maximum likelihood clearly prefers the scale radius of CUO models (see column [4] from Table~\ref{tab: statistics}). This points to a mild preference for a CUO of mass of $939^{+166}_{-331} \, \msun$. 

It is important to mention, though, that such a CUO would have a Plummer projected half-mass radius of $0.016^{+ 0.005}_{- 0.015}$~pc, hence considerably smaller than the values of 0.153~pc and 0.041~pc measured for NGC~3201 and NGC~6397 in Paper~I, respectively. 
This points to a rather concentrated population of remnants, which may be difficult to explain in astrophysical terms. 

\subsubsection{Proper motion dispersion profile}

We also constructed proper motion dispersion profiles to probe the goodness of fit of different dark central mass models. The observed profile was constructed from the data used in our fits, which was composed by \hst\ stars up to $149\farcs2$, and \gaia\ stars beyond this limit. To compute the observed dispersion and its 1-$\sigma$ uncertainties, as well as for other velocity dispersion computations throughout this work, we followed the recipe from \citeauthor{vanderMarel&Anderson&Anderson10}~(\citeyear{vanderMarel&Anderson&Anderson10}, appendix~A). Briefly, the method consists in a maximum likelihood approach that assumes the proper motion spread in each bin to be Gaussian. Next, to correct for the known bias of this method (see \citealt{vandeVen+06}, appendix A), we employ a Monte Carlo approach where we generate $10^{4}$ Gaussian pseudo datasets from the estimated parameters and the observational uncertainties. We then analyse those data in the same fashion as the real one. The statistics of the Monte Carlo results provide both an estimate of the bias in $\sigma_{\mu}$ (which we use to correct our maximum likelihood estimate) and of its uncertainties. 

Figure~\ref{fig: vdisp} displays the computed proper motion dispersion (black circles), according to the recipe explained above, in ten logarithmically-spaced radial bins.\footnote{The edges of the bin are logarithmically spaced, while the black circle is positioned at the mean of the data points.} The horizontal error bars considered the 1-$\sigma$ radial quantization noise, 
\begin{equation}
    \epsilon_{R_{\rm proj}} \equiv \displaystyle{\sqrt{\int_{R_{\rm min}}^{R_{\rm max}} f(R) \left(R - \frac{1}{N} \sum_{i}^{N} R_{i}\right)^{2} {\rm d}R}} \ ,
\end{equation}
where $R_{\rm min}$ and $R_{\rm max}$ are the radial limits of the bin, $R_{i}$ are the data points in the bin, and $f(R)$ is the probability distribution function of projected radii inside the bin, calculated with a KDE approach. Indeed, the $x$-axis error bar should translate our uncertainty on where to place the bin, which is better described by the scatter of radii inside each bin (e.g., figures~2 and 7 from \citealt{Durazo+17} and \citealt{Kacharov+22}, respectively). We notice that, when not accounting for the uncertainties, our modelling with \mpo\ predicts higher proper motion dispersions in the outer regions than observed. 
This excess may be caused by our neglect of mass segregation,\footnote{See table~3 from \cite{Baumgardt+22} for details of mass segregation in M4.} whereas in reality one expects that the average stellar mass decreases with projected distance, and only a 10\% decrease at $\sim 8$ arcmin is sufficient to match model and observations.

In any case, it is the points in the centre that are used to discriminate between models with different mass excess, since the effect of a central dark mass on the velocity dispersion profile is limited to a small influence region (i.e., $R \ll 100\farcs0$).\footnote{When using the relation from \citet{Peebles72}, the radius of influence of a putative IMBH such as the one we fit would be $\sim 14\farcs0$.}
In sum, Figure~\ref{fig: vdisp} supports our previous conclusions: The agreement for projected radii smaller than $10\farcs0$ is clearly poor for the model with no central dark mass, while equally satisfactory for the models with an IMBH or a CUO.
We provide as online material an analogue of Figure~\ref{fig: vdisp} with better resolution, the observed proper motion dispersion we constructed, as well as the fitted profiles for different mass models.

\subsection{Robustness} \label{ssec: robustness}

We now test the robustness of our results under different assumptions. As in Paper~I, we varied the assumptions of cluster centre, replacing our chosen centre with that from \citealt{Goldsbury+10}, lying $0\farcs38$ away from the original position we used; bulk proper motion, replacing ours with that of \citealt{Vasiliev&Baumgardt&Baumgardt21}); and decreasing the maximum allowed proper motions error threshold to half our standard criterion. As seen in lines 9--20, of Table~\ref{tab: results-full}, the best fitted values agree with our diagnostics above within the 1-$\sigma$ error bars, and the AICc diagnostics produce similar conclusions. We also ran mock datasets with 10\% underestimated proper motion uncertainties up to two times the CUO scale radius (as in Paper~I, to probe for unknown systematics) and verified that our results pertaining to the central dark mass did not change significantly.

Thus, the presence of a central mass seems robust, with still mild indications of it being extended. We also verified that the error budget of our data was higher than any imprint of mass segregation in M4, using the same calculations as in Section~5.3.2 of Paper~I. An equivalent of figure~6 of Paper~I is provided as Figure~\ref{fig: err-mag-relation} in the present article.

\begin{figure}
\centering
\includegraphics[width=0.98\hsize]{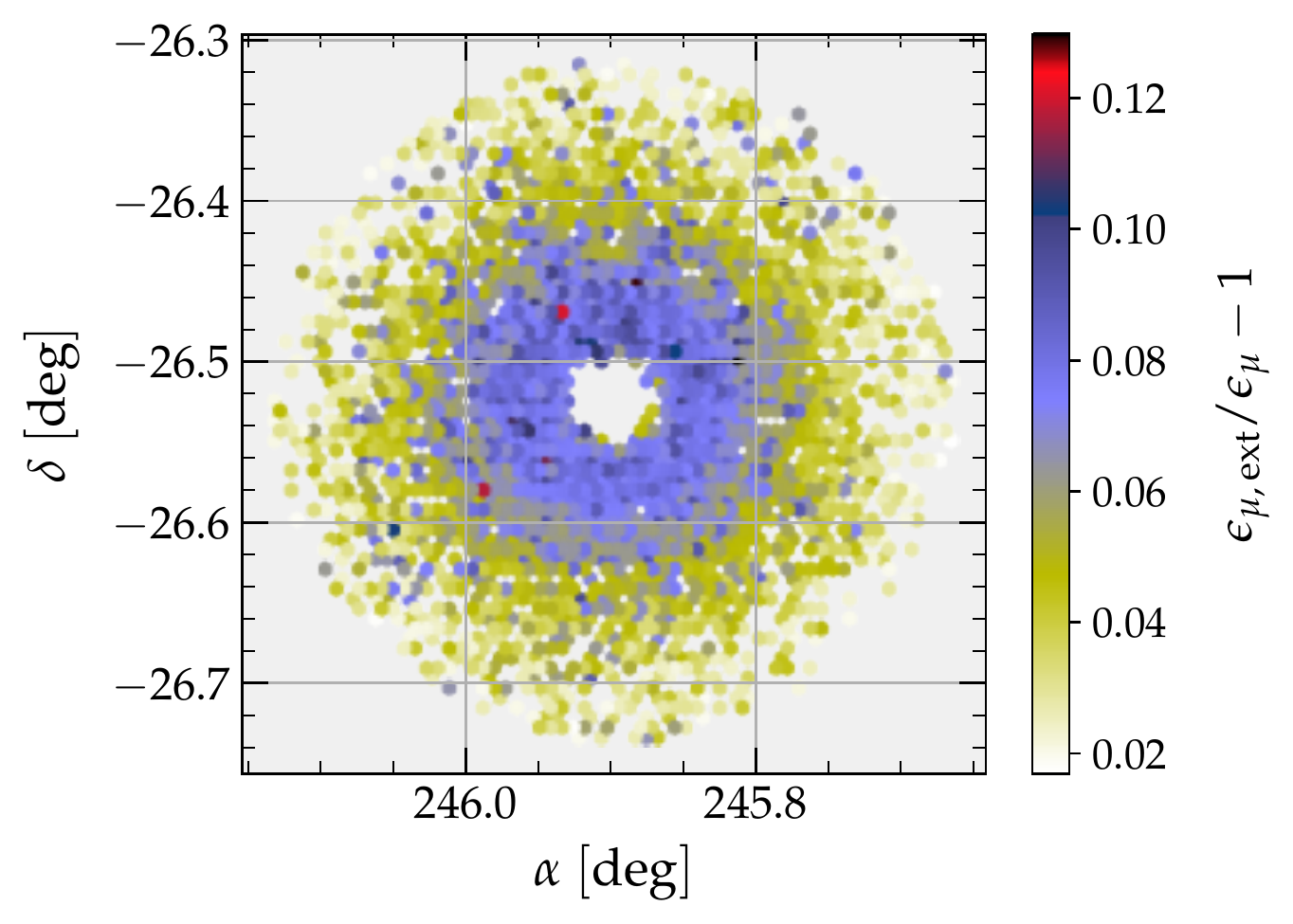}
\caption{\textit{Systematic errors from Gaia:} Underestimation of our adopted  \textit{Gaia} proper motion errors (the statistical errors from the catalogue) by including the systematic errors highlighted by \protect\citeauthor{Vasiliev&Baumgardt&Baumgardt21} (\protect\citeyear{Vasiliev&Baumgardt&Baumgardt21}, see text). The hole in the centre represents the regions where we used \hst\ data. The bulk of the stars, quantified by the 84th percentile, have underestimated proper motion errors (i.e., $\epsilon_{\mu, \rm ext} / \epsilon_{\mu} - 1$) $< 9$ per cent. 
The respective median (50th percentile) is of 6\%. Lines 21--24 from Table~\ref{tab: results-full} display the \mpo\ fits considering an error budget corrected for such systematics (assuming this same scaling factor for both $\epsilon_{\mu \alpha}$, $\epsilon_{\mu \delta}$, but with $\rho_{\mu \alpha \delta}$ unchanged).}
\label{fig: gaia-sys}
\end{figure}

\begin{figure*}
\centering
\includegraphics[width=0.99\hsize]{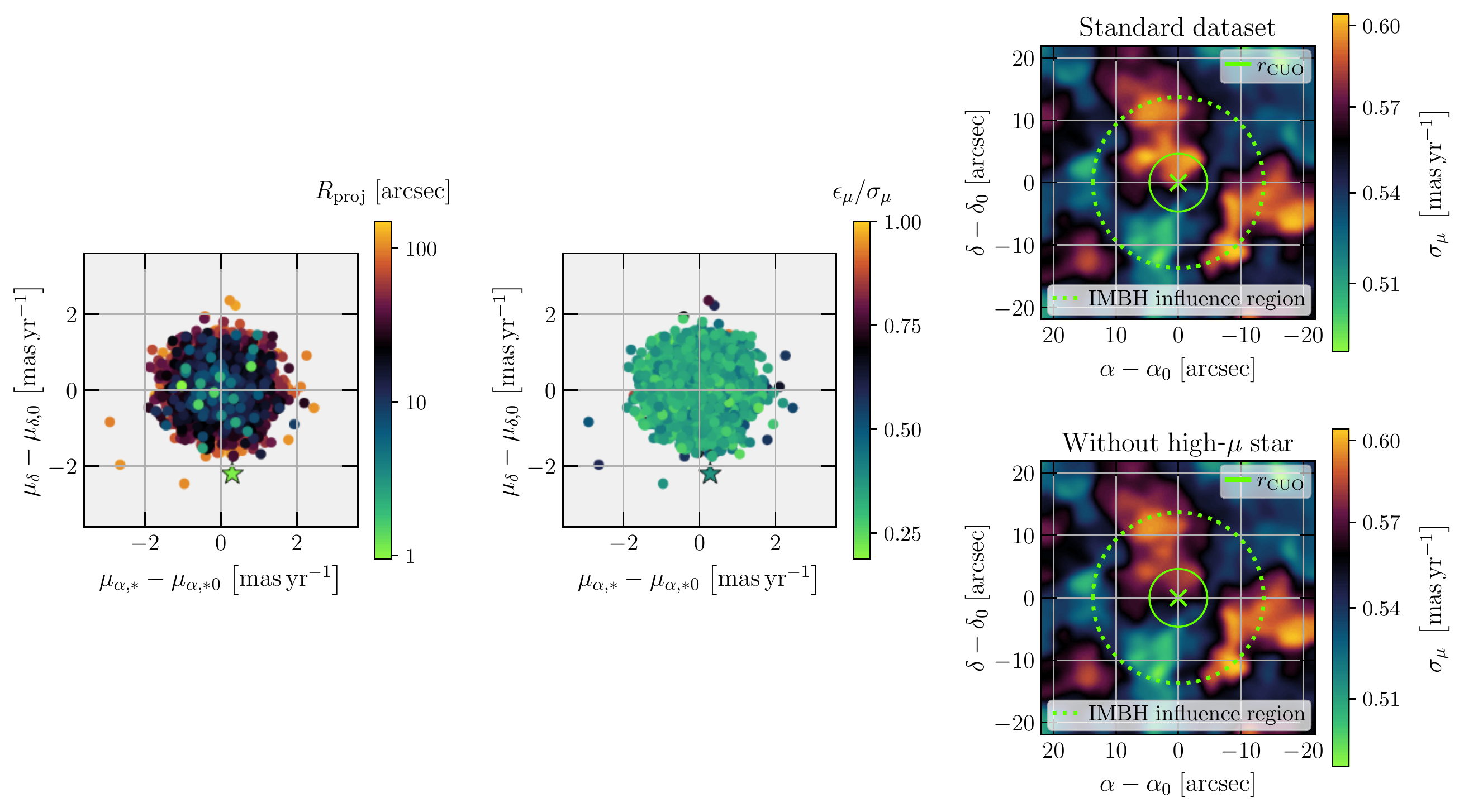}
\caption{{\it High-$\mu$ star:} Different diagnostics related to the high velocity star.
The \textbf{left} and \textbf{middle} columns display the cluster in proper motion space, colour-coded by projected distance to the centre and proper motion error (in units of the local proper motion dispersion), respectively. 
 The proper motion error is defined as in eq.~B2 from \protect\citealt{Lindegren+18}, while the local velocity dispersion is calculated following section~3.3 of Paper~I.
The \emph{asterisk} denotes the high-$\mu$ star.
The \textbf{right} column shows proper motion dispersion maps, constructed according to appendix~B of \protect\citet[][, the specific computation of the velocity disperison per bin followed \protect\citealt{vanderMarel&Anderson&Anderson10}]{Vitral&Boldrini22}, with (\textbf{top}) and without (\textbf{bottom}) the high-$\mu$ star.
The \emph{solid} and \emph{dashed circles} display the scale radius of the sub-cluster of unseen objects and the radius of the sphere of influence of the putative intermediate-mass black hole (Eq.~[\ref{eq: r-imbh}]), respectively.
This figure highlights the method used to spot the high-$\mu$ star, its low error budget and its impact on the velocity dispersion profile.
}
\label{fig: smoking-gun}
\end{figure*}

\subsubsection{\gaia\ systematics}

On top of the previous robustness tests, we added a new one concerning the \gaia\ EDR3 systematics. Indeed, \gaia\ EDR3 data presents an inconvenient issue related to spatially correlated systematic errors \citep[e.g.,][]{Lindegren+21}, which are usually associated with the telescope scan directions. The modelling and correction of these systematics in our data is beyond the scope of this work, and we only use the statistical errors provided in the catalogue. In fact, the impact of these systematics on GCs is not yet very clear, with recent works focusing more on describing them rather than presenting a method to correct for them \citep[e.g.,][]{Fardal+21}.

The most robust correction for these systematics in GCs is perhaps the one given by \cite{Vasiliev&Baumgardt&Baumgardt21}, $\epsilon_{\mu,\rm ext}$, obtained (their eq. 3) by summing in quadrature the statistical errors -- multiplied by a scaling factor dependent on surface density (their table~1) -- and their derived systematic errors  (i.e., $\epsilon_{\mu, \rm sys} = 0.026 \ \masyr$).
As seen in Figure~\ref{fig: gaia-sys}, the bulk of the \gaia\ stars, quantified by the 84th percentile, have underestimated proper motion errors (i.e., $\epsilon_{\mu, \rm ext} / \epsilon_{\mu} - 1$) by only $< 9\%$. However, their individual, separate effects on $\epsilon_{\mu \alpha}$, $\epsilon_{\mu \delta}$ and $\rho_{\mu \alpha \delta}$, which we use in our Jeans modelling, are not yet well quantified. Assuming the same factor for all these components could insert new systematics, which in turn are beyond the scope of our modelling.

Nonetheless, for checking purposes, we constructed the data with this same scaling factor for both $\epsilon_{\mu \alpha}$, $\epsilon_{\mu \delta}$, while keeping $\rho_{\mu \alpha \delta}$ unchanged, and ran our four main mass models with these corrected errors. The results are displayed in lines 21--24 from Table~\ref{tab: results-full}, and the reader can once again see that the results still point strongly to a central dark mass with only mild evidence to be extended
(qualitatively similar AICc diagnostics). Hence, it appears reasonable to neglect these systematics in our modelling.

\begin{figure*}
\centering
\includegraphics[width=0.247\hsize]{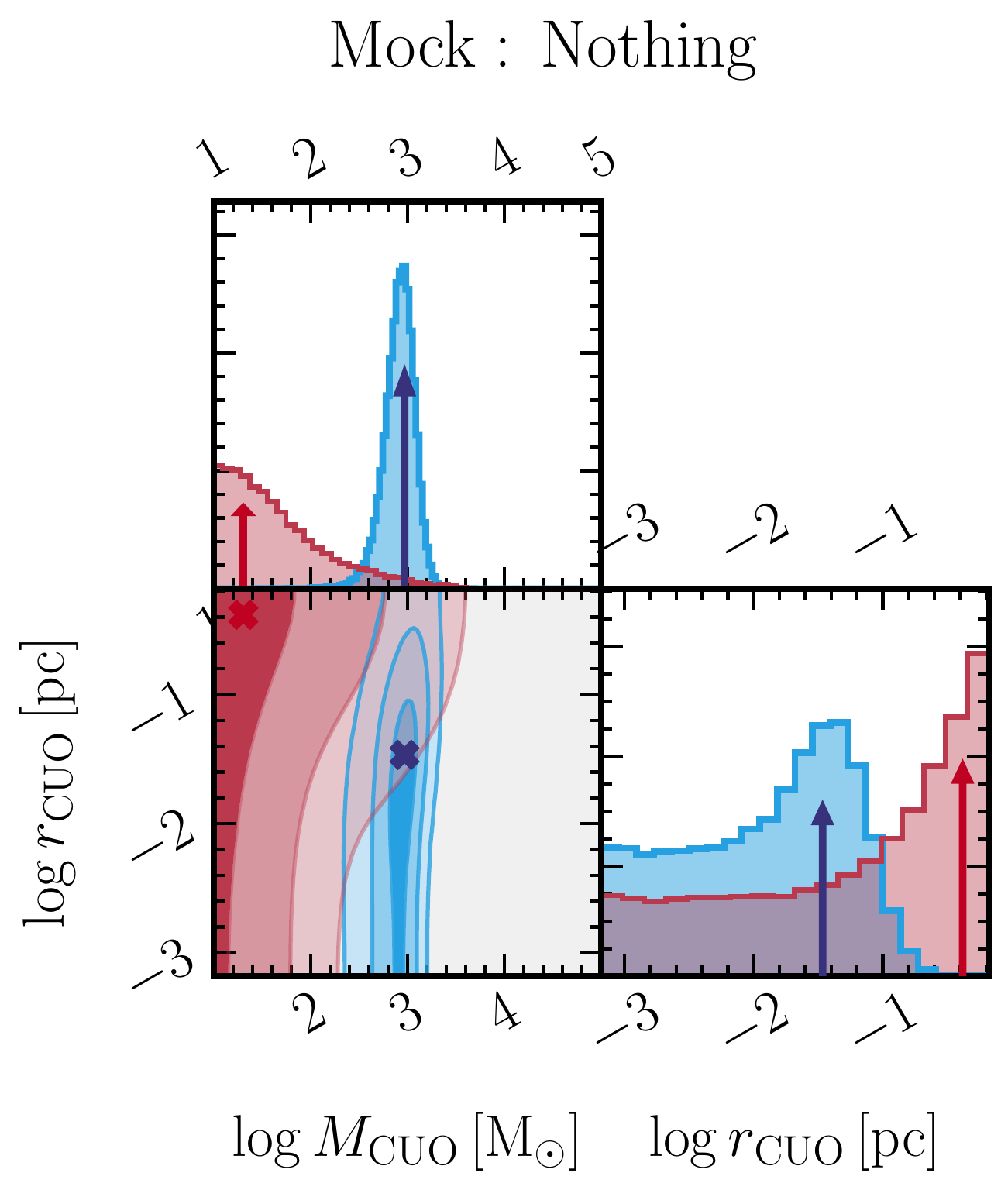}
\includegraphics[width=0.247\hsize]{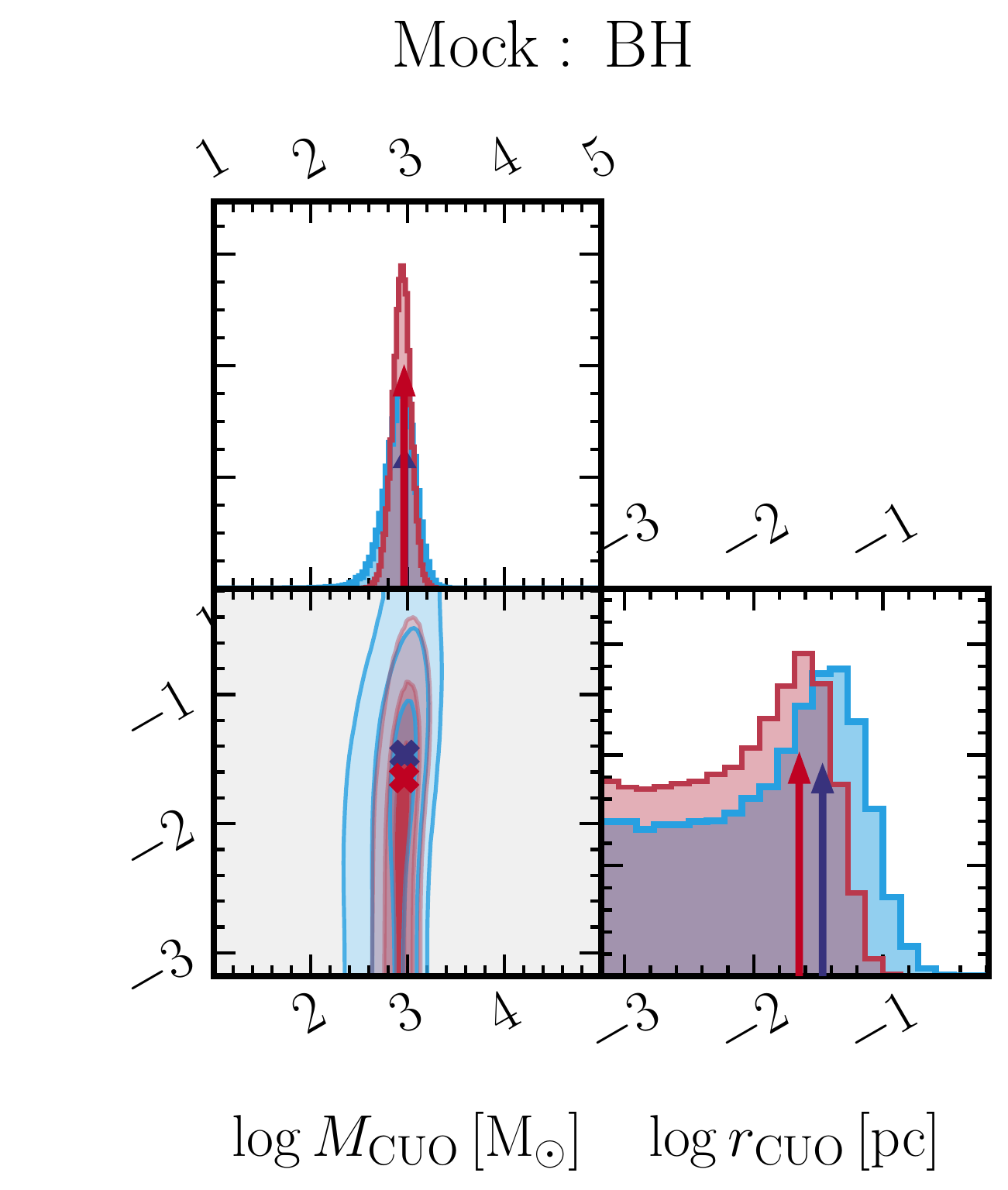}
\includegraphics[width=0.247\hsize]{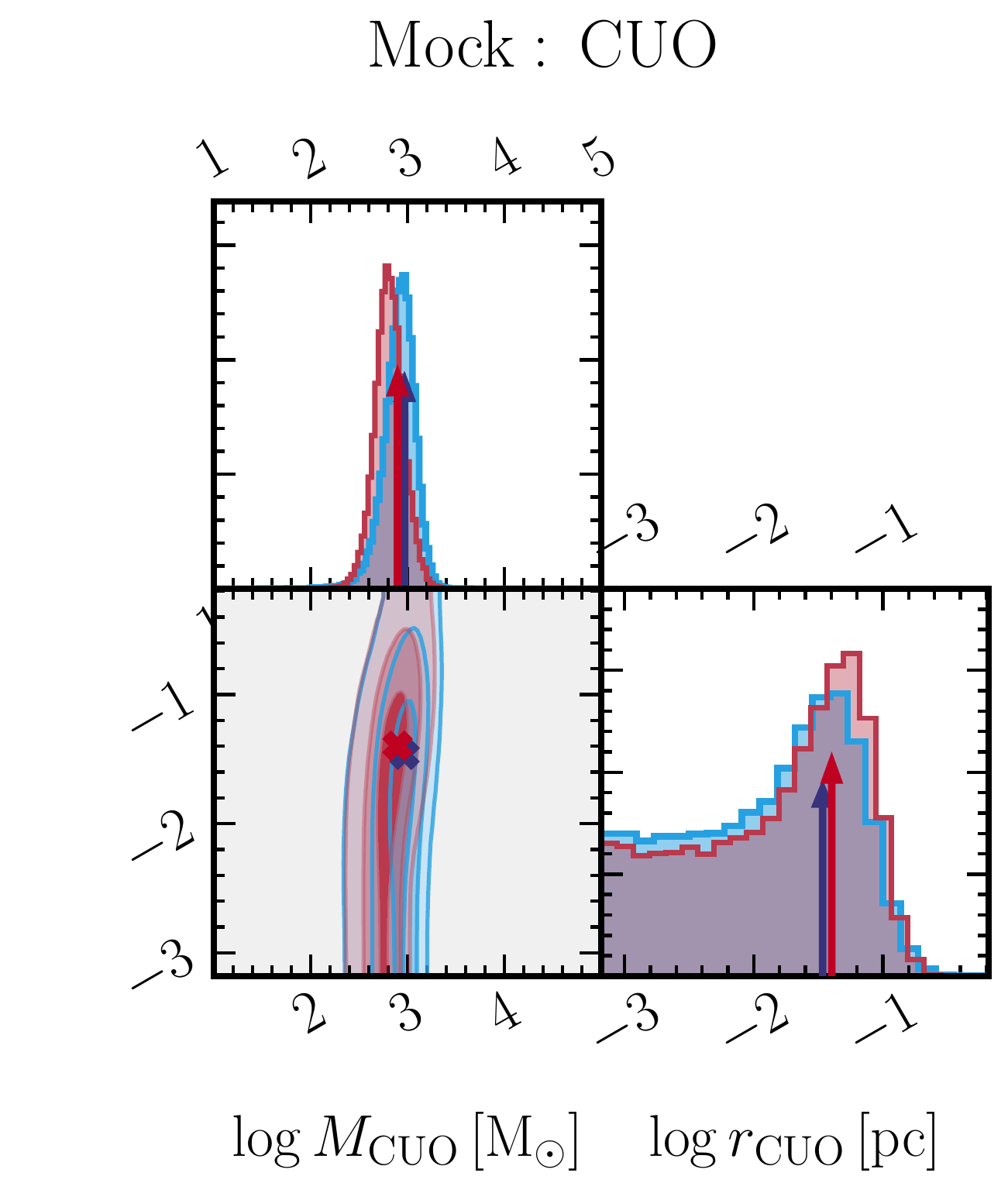}
\includegraphics[width=0.247\hsize]{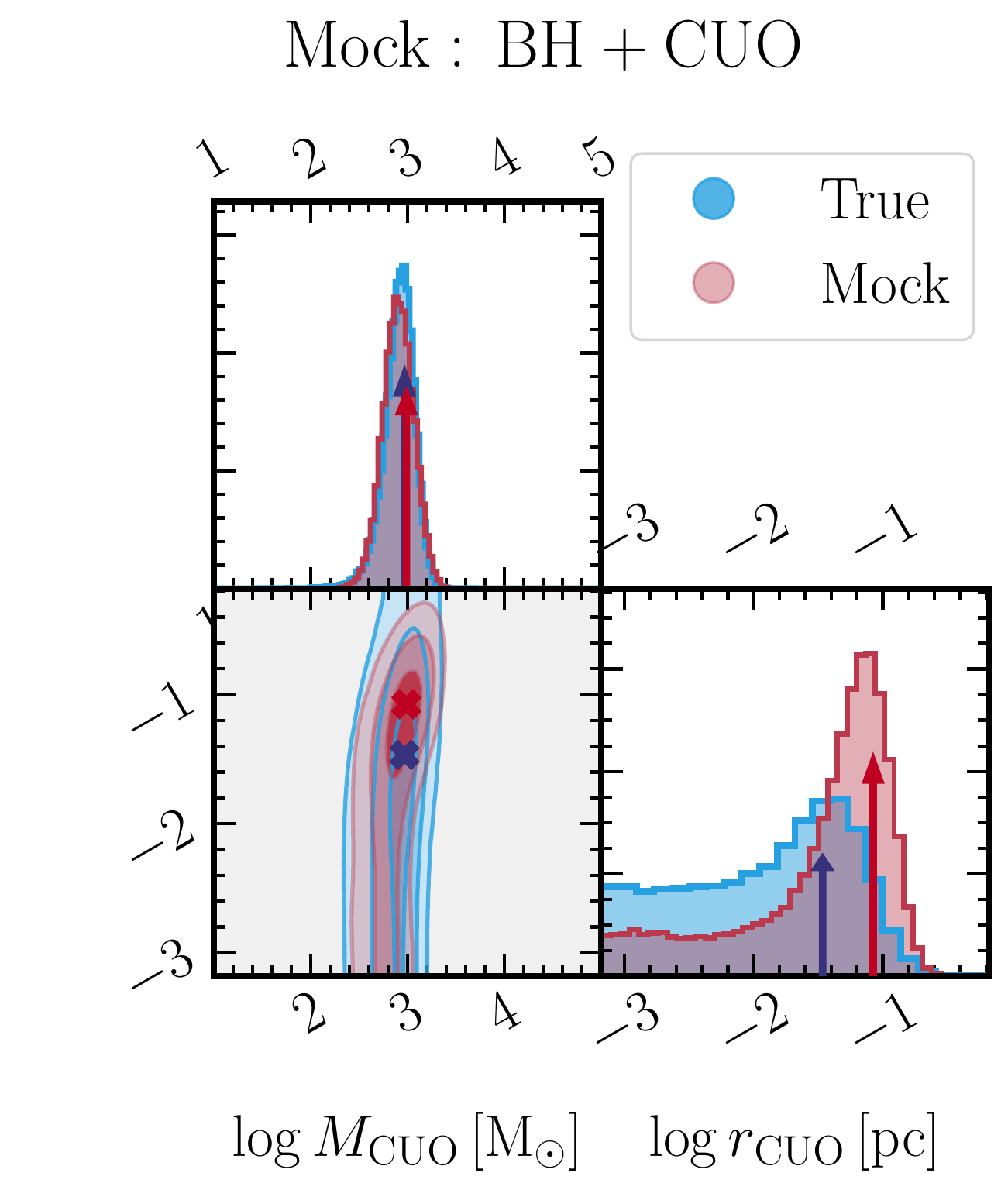}
\caption{{\it Mock data comparison without the high-$\mu$ star:} 
Same as Figure~\ref{fig: comp-standard}, without the \textit{high-$\mu$ star} (see Section~\ref{ssec: high-v}) in \textit{blue} and the mock data (constructed with \agama) in \textit{red}, but now using the values from lines 25--28, on Table~\ref{tab: results-full}.}
\label{fig: comp-highv}
\end{figure*}

\subsection{High velocity stars} \label{ssec: high-v}

Our previous \mpo\ fits and respective robustness tests seem to indicate that although M4 has a central mass excess consistent with a $939^{+166}_{-331} \, \msun$ CUO, the extension of this mass, namely 0.016~pc, still seems too mild to be reliably associated with a collection of dark remnants, leaving open the possibility of a point-like IMBH. If, indeed, M4 has an IMBH at its centre, one can infer its sphere of influence as the radius at which a star can still have its dynamics considerably impacted by such a black hole \citep{Peebles72}:

\begin{equation} \label{eq: r-imbh}
    r_{\rm BH} \equiv \frac{G M_{\rm BH}}{\sigma^2} \ ,
\end{equation}
where $M_{\rm BH}$ is the black hole mass and $\sigma$ the characteristic velocity dispersion at this region. Substituting the values from our fit and data,\footnote{We use $M_{\rm BH}=792 \, \msun$ and $\sigma = 0.6 \, \masyr$, picked for being the 97.7 percentile of the innermost proper motion dispersion computed for M4 in \citealt{Vasiliev&Baumgardt&Baumgardt21}.} one has $r_{\rm BH} = 14\farcs$ Hence, a reasonable check is to see if our data has any high velocity star inside this radius. A high velocity star is defined here as a star having a proper motion modulus beyond a few times the local cluster's velocity dispersion.

In the leftmost panel of Figure~\ref{fig: smoking-gun}, we see the scattered distribution of \hst\ stars in proper motion space, colour-coded according to their distance to the cluster's centre. Indeed, one can clearly observe that at roughly ($\mu_{\alpha,*}$$-$$\mu_{\alpha,*0}$, $\mu_{\delta}$$-$$\mu_{\delta}$) $\approx$ ($0.3,-2.2$), there is a star located at roughly $1\farcs07$ from the GC centre.
The proper motion offset of this star corresponds to $3.7$ times the local velocity dispersion.
We highlight this star with the $\star$ symbol in the plot, and label it hereafter as the \emph{high-$\mu$ star}.\footnote{For context, this star has a F606W magnitude of $21.01 \pm 0.05$.}
One can also see in the middle panel that the high-$\mu$ star presents a reasonably low proper motion error, below half of the local velocity dispersion.

\subsubsection{Assessing the influence of the high-$\mu$ star}

After discovering this high velocity star, we tested how much our results depend on it. Since \mpo\ is a Bayesian code, each star contributes separately to the model's likelihood. Therefore, parameters related to the inner cluster such as central mass excess and dark mass scale radius are sensitive to the innermost stars.
Indeed, as seen in the right panels of Figure~\ref{fig: smoking-gun}, the velocity dispersion map inside the sphere of influence of the eventual IMBH considerably decreases after the removal of the high-$\mu$ star (see region right above the $\times$ symbol), which can affect the fits of the parameters pertaining to this central region. Thus, we decided to run \mpo\ again, with the same subset as the standard one, but without this high-$\mu$ star.

The results are listed in lines 25--28 of Table~\ref{tab: results-full}. One notices that we lose just a negligible AICc evidence in favour of a central mass, with now a $0.2\%$ probability of no mass excess whatsoever in M4. Although the IMBH model is preferred by AICc, the differences in AICc for models with a mass excess are too low to distinguish the IMBH and CUO scenarios. 

We thus turn once again to comparing the CUO constraints that \mpo\ obtains from the data with those it obtains on mock data.
Figure~\ref{fig: comp-highv} displays this comparison, in a similar fashion as Figure~\ref{fig: comp-standard}, with statistical tests provided in Table~\ref{tab: statistics} (Test ID: $v_{\star}$). We still obtain strong confidence for it when comparing the CUO marginal distributions obtained on the observed data and on the mock with no central mass excess (Nothing). Indeed, a lack of excess mass would shift the respective marginal distribution towards much lower values than observed in our real data.

For this subset without the high-$\mu$ star, one notices two fundamental differences in the comparison of the marginal CUO radius distributions with the analogous comparison for the dataset including the high-$\mu$ star (Fig.~\ref{fig: comp-standard}): 1) The \mpo\ fits of observed data yield a higher scale radius for the dark mass 2) Even the mock with an IMBH yields an extended population, likely due to completeness issues (see section~\ref{ssec: diagnosis-mu}).
Indeed, by removing a very inner star with high proper motion, it is reasonable to assume that \mpo\ will interpret the new data as having a more diffuse mass excess, rather than concentrating it towards the centre.
On the other hand, the outer velocity anisotropy remains significantly tangential. In summary, the removal of the high-$\mu$ star does not affect our diagnosis of a central mass of roughly $800 \, \msun$ (see rows~26 and 27 of Table~\ref{tab: results-full}),
but renders the choice of mass model even more complicated, as \mpo\ runs on the mocks with IMBH or with CUO yield similar marginal distributions to the fits on observed data (as confirmed statistically in Table~\ref{tab: statistics}), on top of providing a more realistic scale radius for the CUO component (now roughly double in size).

\section{Discussion} \label{sec: discussion}

\subsection{Tangential outer anisotropy} \label{ssec: tangential}

Our measurement of inner isotropy and outer tangential anisotropy in M4 has many interesting implications.
Recently, \cite{Aros+20} analysed different Monte Carlo mass models to probe the impact of an IMBH or a CUO on the velocity anisotropy of GCs, but found no scenario with outer tangential anisotropy. 
Indeed, many studies found it difficult to reproduce tangential orbits such as in our Figure~\ref{fig: anis-mpo} \citep{Oh&Lin92,Vesperini+14,Tiongco+16,Zocchi+16}. 

However, \cite*{Bianchini+17} measured outer tangential orbits in simulations of clusters evolving in a tidal field (see their Figure~3), and related them to GCs that have suffered stronger tidal interactions.
They argued that tidal interactions in the cluster outskirts tend to prune stars on radial orbits more severely than the ones on tangential orbits.  This pruning shortens the relaxation time of the cluster and increases the total mass loss.\footnote{\cite{Bianchini+17} suggests that tangential anisotropy is related to relaxation times $\lesssim 10^9$~Gyr and mass losses $\gtrsim 60\%$.}  The stronger tidal field in the cluster outskirts also impacts low mass stars, which tend to lie in the outer regions of the cluster, resulting in a higher mean stellar mass for the remaining system.  
The same tendency for outer tangential orbits applies for clusters with stronger mass segregation, hence where stellar encounters heated more low-mass stars, pushing them on sufficiently elongated orbits to escape the cluster with the help of tidal forces from the Milky Way.

Furthermore, \cite{Baumgardt&Makino&Makino03} used simulations to show that the amount of tangential anisotropy remains more or less constant until near core-collapse, at which point it starts to decrease. This is because more stars are scattered out of the core on radial orbits into the cluster outskirts after it reaches core-collapse, as a result of higher inner densities hence more stellar encounters and corresponding exchanges of energy.

All in all, the  tangentially anisotropic outer profile of M4 seems to indicate that it has suffered very strong tidal interactions. This is consistent with its very low pericentre ($0.4-0.6\ \rm kpc$, \citealt{GaiaHelmi+18,Sun+22}), as compared with $2.5-2.9$~kpc and $8.4-8.5$~kpc for NGC~6397 and NGC~3201 respectively, where the orbits measured in Paper~I were consistent with isotropy. 
As a result, M4's mean stellar mass should be higher than expected for an isolated system. 
Furthermore, its
total mass should have been considerably higher in the past. 
The tangential outer anisotropy profile also suggests that M4 has not yet reached core-collapse (as argued in Section~\ref{ssec: overview-m4}). 

\subsection{Diagnosis of the high-\texorpdfstring{$\mu$}{PM} star} \label{ssec: diagnosis-mu}

Although the central mass excess in M4 is robust to the removal of the high-$\mu$ star, the extension of this mass becomes significantly larger. Hence, we present below a statistical assessment on the probability of finding such a high proper-motion star, an analysis backward in time of the star's proper motion vector to probe its association with a putative IMBH, and finally an analysis of the reliability of the the high-$\mu$ star parameters inferred from \hst.

%
\subsubsection{Statistical diagnosis}

One way to infer how likely is it to measure a high-$\mu$ star in M4 is to compute the probability of finding such high velocity stars in a mock dataset that mimics the data.  To this end, we define a high-$\mu$ star as a star inside the eventual IMBH influence region, having a proper motion modulus $\mu$ higher than $n$ times the local proper motion dispersion ($\sigma$, taken as in Eq.~[\ref{eq: r-imbh}]), where we set $n$ as the geometric mean of the highest and second highest $n_i$ in the data,\footnote{The highest $n_i$ (that of the high-$\mu$ star) is $3.71$, and the second highest is $3.57$.} which yields $n = 3.64$. 
This is motivated by the fact that in our subset, the highest $n_i$ (i.e., that of the high-$\mu$ star) has an important impact on the extent of the central mass, while the second highest does not (see Figure~\ref{fig: comp-highv}). Next, we follow the recipe below:

\begin{enumerate}
    \item We construct a mock dataset with isotropic velocities, as we did in Sect.~\ref{ssec: dark-mass}, with an IMBH of $792\, \msun$. 
    \item We consider only stars inside the influence region of this black hole, taken to be  $14\farcs$
    \item We store the number of stars such that $\mu > n \,\sigma$. 
    \item We repeat these procedures 1000 times.
\end{enumerate}

This computation delivers a 56\% probability of having no high-$\mu$ star in the subset, with the probabilities of having $N$ high-$\mu$ stars being 33\%, 9\% and 2\% for $N = 1$, 2 and 3, respectively. Such numbers are reassuring since our case falls in a 33\% probability scenario, which is having a single high-$\mu$ star, thus not an unlikely one (i.e., as much as half as likely than the case with no high-$\mu$ star).
Hence, we conclude that if M4 has an IMBH with the same characteristics from our fit, the existence of a single high-$\mu$ star in our final subset, as is the case, is statistically appropriate. 

\begin{figure}
\centering
\includegraphics[width=0.9\hsize]{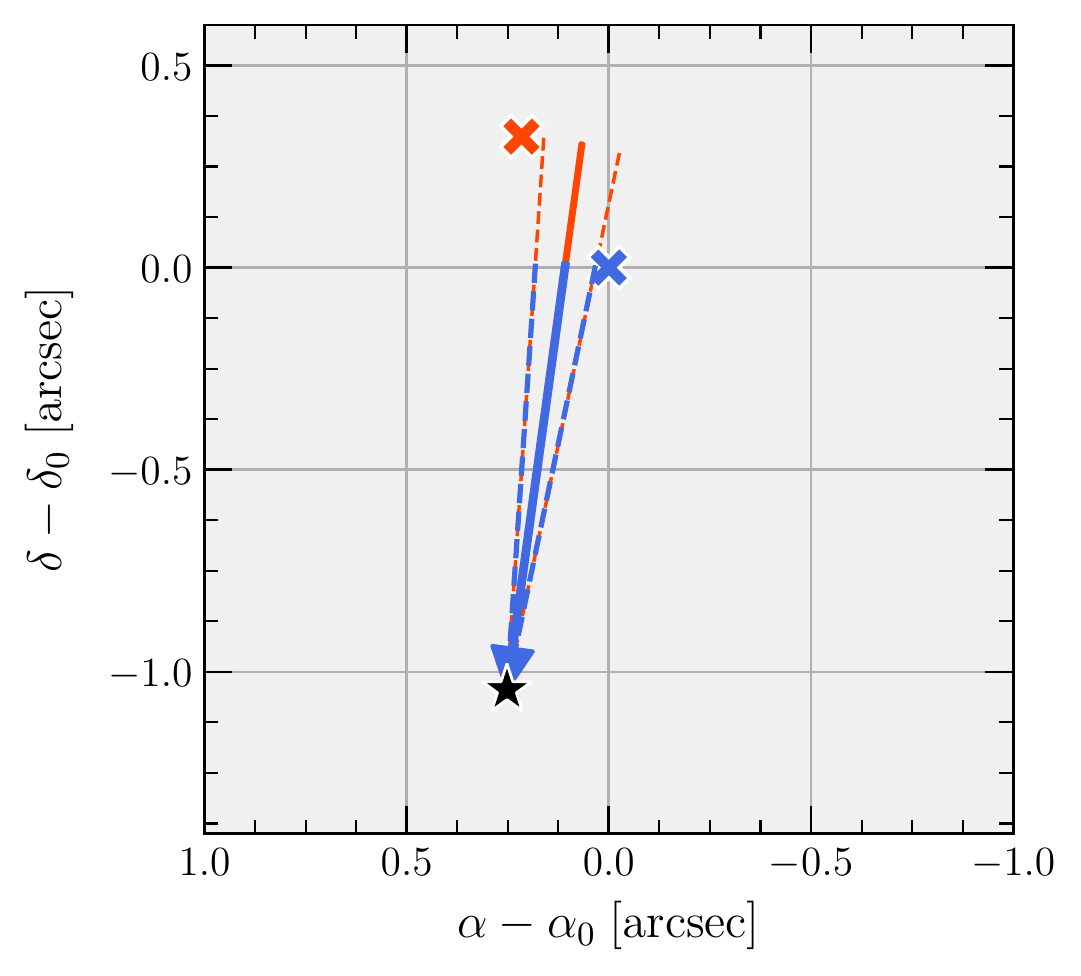}
\caption{\textit{Vector analysis:} Sky projection of the proper motion vector of the high-$\mu$ star with respect to its closest passage to the cluster's centre. The high-$\mu$ star is shown in black, while the centres from \protect\cite{Vitral21} and \protect\cite{Goldsbury+10} are shown as \textit{blue} and \textit{red} crosses, respectively. The dashed lines represent the 1-$\sigma$ uncertainty on the proper motion vector, derived from a Monte Carlo approach considering Gaussian uncertainties. This plot shows that the high-$\mu$ star could in principle be coming from within $0\farcs1$ of the cluster's centre.}
\label{fig: pm-direction}
\end{figure}

\subsubsection{Dynamical past}

If the high-$\mu$ star is associated with a putative IMBH, its high velocity\footnote{For context, the high-$\mu$ star has a velocity of $19.5 \pm 1.4 \, \kms$, and the escape velocity in the centre of M4 (using the values form our fits) is $21.8 \, \kms$.} could be eventually explained by a dynamical kick after close passage to the compact object \cite[e.g.,][]{Hills88}. To evaluate the likeliness of this scenario, we followed the recipe from \citet[][section 5.1.2]{Libralato+21} by verifying that the high-$\mu$ was a good candidate to be coming radially from M4's centre. As in the previous work, we considered it as a good candidate if:


\begin{enumerate}
    \item The angle $\theta$ between the proper motion direction and the direction from the cluster's centre to the star was of less than $10^\circ$ (this value is taken as the same from \citealt{Libralato+21}, for consistency).
    \item The closest distance to the cluster centre backward in time, based on the relative proper motion vector
    ($\pm$$1\,\sigma$), is smaller than the radius of the sphere of influence of the IMBH, $14\farcs$ (see Sect.~\ref{ssec: high-v}).
    \item The closest approach occurred within the age of the star and the putative black hole. Since black hole formation through runaway mergers is thought to occur in the earliest phases of the cluster dynamical evolution (see figure~1 from \citealt{Gonzalez+21}), and the high-$\mu$ star is located on the main sequence, this age should be of the order of $\sim 10$ Gyr.
\end{enumerate}

We employed a Monte Carlo approach to estimate the 1-$\sigma$ uncertainty on the proper motion direction, where we perturbed a thousand times the proper motion by a Gaussian noise of dispersion equivalent to the proper motion error, as in \cite{Libralato+21}. The result is displayed in Figure~\ref{fig: pm-direction}, with the colour blue denoting the calculations with the standard centre from \cite{Vitral21}, while the calculations with the centre from \cite{Goldsbury+10} are shown in red.  In both cases, the proper motion is aligned with the direction to the cluster centre by less than 6$^\circ$, and the closest passage of the high-$\mu$ star is less than 478 or 612 years ago for the respective centres of \cite{Vitral21} or \cite{Goldsbury+10}, thus well below our thresholds. The distance from both centres at this stage is of only $0\farcs1$ (i.e., 0.0009~pc), much smaller than the threshold we set.

This test helps to confirm the high-$\mu$ star as a potential and interesting target for future follow-up studies, but by no means proves the existence of an IMBH. Indeed, the analysis is purely based on proper motions, and the line-of-sight component from the velocity vector of this star could be misaligned with the centre. Moreover, the precise distance of the star to the centre is also unknown, and it could be a projection effect from the cluster's outskirts.

\subsubsection{HST diagnosis}

\begin{figure*}
\centering
\includegraphics[width=0.8\hsize]{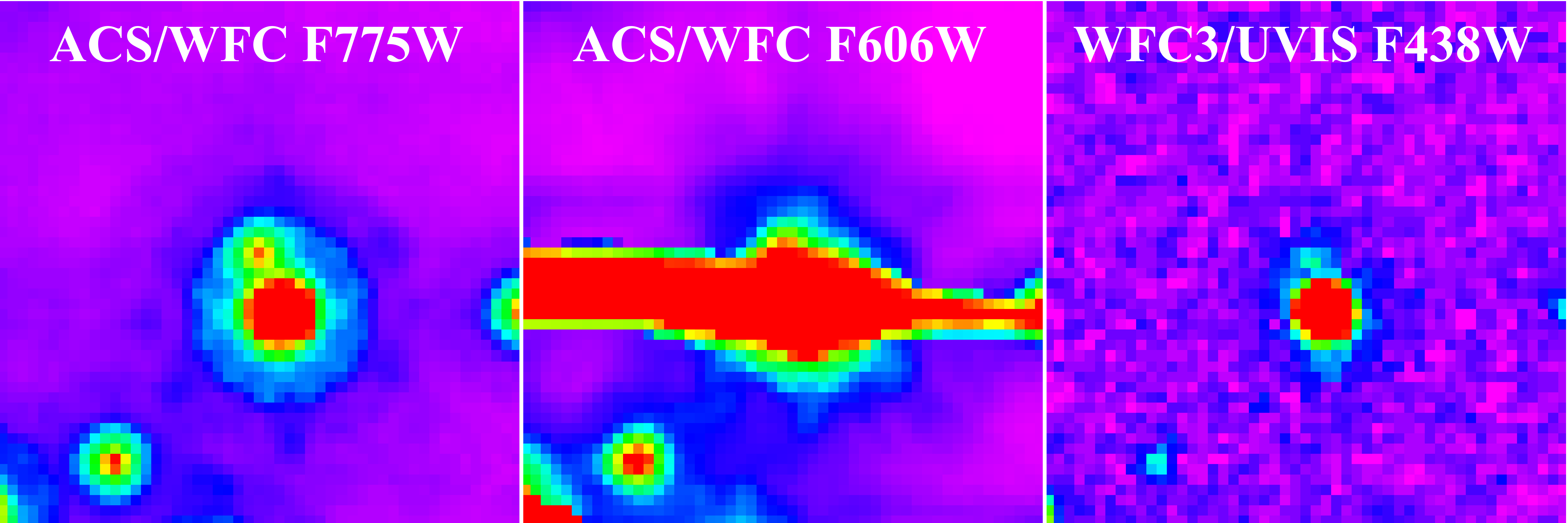}
\caption{\emph{High-$\mu$ star and bright neighbour:} 
The panels display a collection of zoomed-in \hst\ stacked images (logarithmic scale; pixel scale 40~mas$\,$pixel$^{-1}$) in three filters (\textit{left}: ACS/WFC F775W; \text{middle}: ACS/WFC F606W; \textit{right}: WFC3/UVIS F438W) with the high-$\mu$ star (discussed in the text) and its bright nearby neighbour in the centre. 
The \textit{left} panel shows an example where both the high-$\mu$ star and its brighter neighbor are bright and unsaturated. The \textit{middle} panel presents the case where the neighbour star is heavily saturated. 
Finally, the \textit{right} panel highlights the case where the high-$\mu$ star is relatively faint with respect to its neighbour.
}
\label{fig: sg-bright}
\end{figure*}

\begin{figure*}
\centering
\includegraphics[width=0.95\hsize]{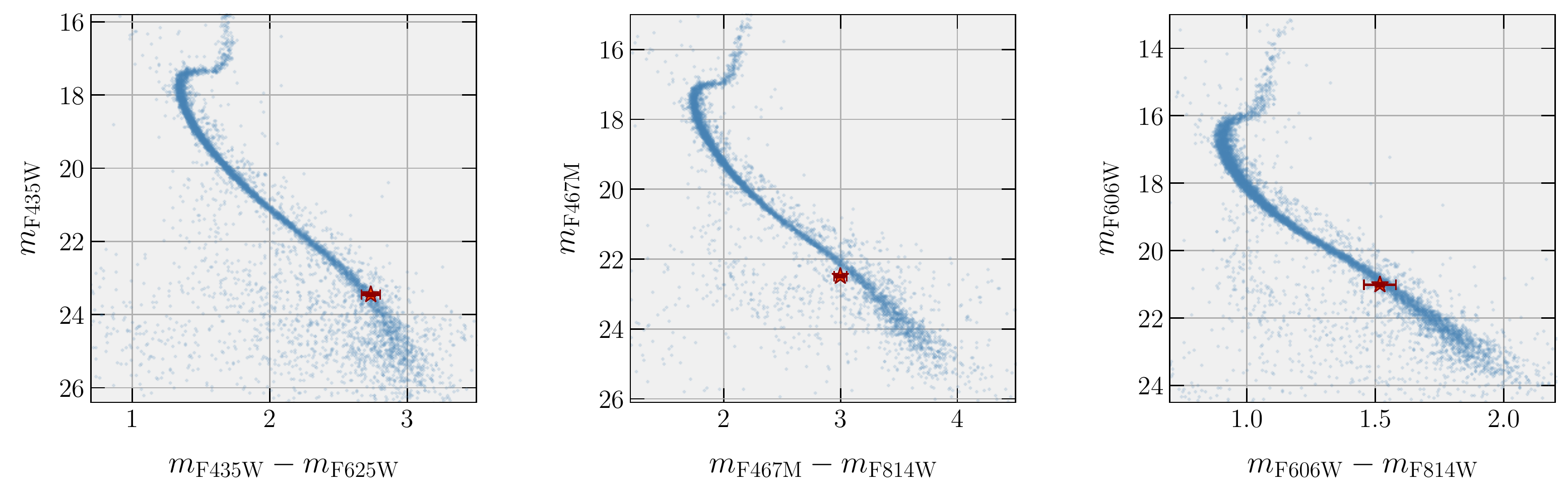}
\caption{\emph{High-$\mu$ star in colour-magnitude diagrams:} The panels display the colour-magnitude diagram of M4 plus a few interlopers, for different selections of filters. Depending on the filter combination, and accounting for the magnitude uncertainties, the \textit{high-$\mu$ star} (red star) is either on the main-sequence (\textbf{right}), or slightly on the blue side next to the main-sequence (\textbf{middle}) or, finally, close to the location of the main-sequence binaries (\textbf{left}). The $X$-axis and $Y$-axis positions (and respective error bars) of the \textit{high-$\mu$ star} in the plots, from left to right, are $[ 2.73 \pm 0.07 , \,  23.46 \pm 0.04 ]$, $[ 3.0 \pm 0.05 , \,  22.49 \pm 0.04 ]$ and $[ 1.52 \pm 0.06 , \,  21.01 \pm 0.05 ]$, respectively. This plots uses data from \protect\cite{Piotto+15,Libralato+22}. 
}
\label{fig: sg-cmd}
\end{figure*}

The high-$\mu$ star is located at ($\alpha, \, \delta$) $=$ ($245^\circ_{.}89676, \,-26^\circ_{.}52613$). 
The photometric catalogues derived in the various steps of the data reduction were obtained with the  \texttt{KS2} software code (\citealt{Bellini+17}; \citealt{Libralato+18,Libralato+19,Libralato+22}). \texttt{KS2} measures position and flux of a star after all its neighbours are point-spread-function (PSF) subtracted from the image. In addition, \texttt{KS2} produces stacked images for each dataset used in the process. We visualised our target object in all these images (Figure~\ref{fig: sg-bright} displays the hi-mu star in various wavebands). 
Our high-$\mu$ star is very close ($0\farcs23$) to a much brighter object (by $\approx 3.4$ magnitudes). Although \texttt{KS2} subtracts all nearby stars prior to estimate position and flux of an object, there can still be some residual contamination that can result in an artificial positional offset. The centroid would be displaced in different ways depending on the camera/filter, and so it could mimic an artificially higher proper motion when multi-epoch images are combined.

We decided to consider a few extra cleaning criteria to assess the reliability of this star. Those criteria can be stricter than our standard ones (and the ones used in Paper~I) since they require the star to pass multiple photometric quality selections on top of the proper motion selections. They were also designed for the outskirts of GCs because the quality thresholds defined in less crowded regions are often more trustworthy. By doing this the other way around (i.e., defining the thresholds from more crowded regions), one could systematically include bad measured stars in the outskirts given worse thresholds near the cluster's core.
In sum, we checked the following points: (i) quality of the photometric fit; (ii) quality of the proper motion fit and (iii) position in the CMD using other filters than F606W and F814W. The diagnosis of each point is given below.

\begin{enumerate}
    \item We initially checked the goodness of the target in our astro-photometric catalogues. Specifically, we looked at the quality of the PSF fit (\texttt{QFIT}), the magnitude rms, the excess/defect of flux outside the core of the star (\texttt{RADXS}) and the fractional flux within the fitting radius prior to neighbour subtraction (\texttt{o} parameter). The high-$\mu$ star was often poorly measured in the analysed filters, sometimes because the quality indicators are considerably beyond our the thresholds we set, sometimes because they are slightly worse than those thresholds. Hence, the star does not pass our new photometric criteria.
  
    \item We then checked the proper motion fit per~se. The original fit was visually inspected and we found no obvious evidence of a problematic fit. We re-ran the proper motion code removing (1) all images in which the nearby neighbour is saturated (as in the middle panel of Fig.~\ref{fig: sg-bright}),
    and (2) all images in which the high-$\mu$ star is very faint (as in the right panel of Fig.~\ref{fig: sg-bright}).
    Regardless, the results of the fits are still consistent within their respective 1-$\sigma$ uncertainty 
    to the proper motion obtained using all images. Hence, the proper motion measurement of this star seems robust to the new criteria.
    
    \item Figure~\ref{fig: sg-cmd} displays three CMDs in instrumental magnitudes with the position of the high-$\mu$ star highlighted (all stars are shown, not only M4 members). Depending on the filter combination, the target is either on the main-sequence (right panel), or on the blue side of the main-sequence (middle panel) or finally, slightly on the location of the main-sequence binaries\footnote{If the high-$\mu$ star is part of a tight non-resolved binary, one could indeed expect an enhancement of its velocity due to binary motions.} (left panel).
\end{enumerate}

These tests shed some doubts on the reliability of the high proper motion of the high-$\mu$ star.
The proper motion of the star is $\sim 2 \, \masyr$, which corresponds to $\sim 0.05$ WFC3/UVIS pixel$\,$yr$^{-1}$. A non-perfect PSF subtraction of the neighbour objects could easily create an uncertainty on the target centroid this large. While the data reduction discussed in \cite{Libralato+22} is specifically designed to deal with crowded environments, it is still hard to interpret a single object. Thus, we believe that the interpretation of this high-$\mu$ star would benefit of additional follow ups. It does not mean that the high-$\mu$ star must necessarily be put apart from our standard subset used for mass modelling, but rather that its influence in the extension of the central dark mass should be interpreted with caution.

Finally, given the dependence of the CUO parameters fitted by \mpo\ on this high-$\mu$ star, we asked ourselves if other central stars failing our new stricter criteria could have biased our results so far. We identified four extra stars that were located up to twice our fitted CUO scale radius and also did not pass such criteria and ran \mpo\ on a subset excluding them, to probe their influence. The marginal distributions of the CUO parameters in this new subset remained nearly identical to the case where just the high-$\mu$ star was removed, meaning that significant changes on our fits of CUO parameters depended rather on the high-$\mu$ star than on our standard cleaning criteria.

To increase the reliability of such our dataset, the two better options would be (1) to increase our subset (hence the number of measurable high-$\mu$ stars), which can be made by having more epochs, using new pointings with \hst\ and/or the \textit{James Webb Space Telescope} \citep[such as in][]{Libralato+23}, thus increasing the quality of our measurements and likely removing less stars in our filtering routine; (2) to perform a follow up study of this particular high-$\mu$ star, again by means of new observations.
These would be necessary steps to further confirm our fits by similar methods.


\subsection{Formation \& retention of an IMBH} \label{ssec: imbh-retention}

To better interpret our fit mimicking an IMBH signature, it is important to understand how such a source could have been formed, and whether its retention is feasible. 
IMBHs may form in GCs via successive black hole mergers over the lifetime of the cluster or via runaway stellar mergers at very early times. 
In the former scenario, initially proposed by \cite{Miller&Hamilton02}, dynamical friction causes the most massive stellar-mass black holes ($\gtrsim 40 \, \msun$) 
to sink to the centre of the gravitational potential well, followed by growth in mass through mergers with other black holes \citep[e.g.,][]{Miller&Hamilton02_4body}
as well as with other typically massive stars \citep[e.g.,][]{Giersz+15}. Such a scenario requires however a very deep potential well 
to avoid ejecting the black hole as it conserves linear momentum with the anisotropic gravitational waves emitted after it merges with other black holes \citep{Peres62,Lousto+10}.

The stellar runaway scenario seems more plausible for a cluster like M4.
In this scenario, an initially massive star suffers multiple physical collisions with other stars during the first few Myr of the GC, before the stars collapse into compact objects.  While the classic runaway model \citep{PortegiesZwart&McMillan02,PortegiesZwart+04} suggests that most of the massive stars merge, potentially collapsing directly into a massive IMBH of $\gtrsim10^4 \, \msun$, an alternative scenario scenario recently 
explored by \cite{DiCarlo2020}, \cite{Kremer2020} and \cite{Gonzalez+21} 
argues that, in some cases, only a handful of massive stars merge, thus forming a small IMBH (i.e., a few hundred solar masses), consistent with the values we fit,\footnote{The range of IMBH masses relative to the 5th and 95th percentiles of our IMBH model (i.e., model~6 from Table~\ref{tab: results-full}) is $441 - 1210 \, \msun$.} in addition to the usual population of stellar-mass black holes. 

An important caveat is that the 
details of this runaway process are highly uncertain. The total mass of the stellar collision products (roughly a few hundred solar masses or more) suggests that IMBH formation is possible, 
assuming that the entire star can directly collapse to a black hole of comparable mass. This assumption requires that little mass is lost: (i) 
dynamically during the collisions themselves, (ii) 
through stellar winds of the stellar collision products prior to collapse, and (iii) 
as stellar ejecta and/or neutrinos during the final collapse to a black hole. Recent studies \citep{Ballone+2022,Costa+2022} suggest that these assumptions may apply in some contexts. However in general, it is considerably uncertain whether IMBH formation is the outcome at early epochs.
 
Even if an $800 \, \msun$ IMBH is formed via stellar collisions while the cluster is very young, one may wonder if it would be retained in the long term. The IMBH will likely be accompanied by a much a larger population of normal stellar-mass black holes with masses $\sim20-40 \, \msun$. Most surely, the IMBH will merge with these stellar-mass black holes and receive a gravitational wave recoil kick,\footnote{An interesting consequence of such recoil kicks would be the possibility of an off-centre IMBH, but we do not explore this scenario due to limitations pertaining to the Jeans modelling assumptions.} as mentioned above \citep[e.g.,][]{GonzalezPrieto+22}. These kicks are reduced due to the high mass ratios, but still may be large enough to eject the IMBH from the host cluster. We can estimate the escape velocity from M4's centre, $v_{\rm esc} = \sqrt{-2\,\Phi(0)}$, for our best IMBH model (i.e., line 6 from Table~\ref{tab: results-full}), given the central potential $\Phi_0 = -(2/\pi)\,b^n\,\Gamma(n)/\Gamma(2n)\,G\,M/R_{\rm e}$ for the S\'ersic model \citep{Ciotti91}, where $n$ is the S\'ersic index, $R_{\rm e}$ the effective radius (see footnote~\ref{ftn:Re}), and $b(n)$ follows the relation given in \cite{Ciotti&Bertin99}. This yields $v_{\rm esc} = 22 \, \kms$, which remains of the same order of typical recoil kick velocities from non-spinning merging black holes with mass ratios of $\sim 20$ \citep[e.g.,][]{Merritt+04_recoilkick,Campanelli+07,Schnittman&Buonanno&Buonanno07,LeTiec+10,Lousto+10,Gerosa+18}. However, if M4 was much more massive at first,\footnote{See \cite{Atallah+22} for a similar analysis taking into account more massive nuclear star clusters, in galactic nuclei.} because of severe mass loss from the tides exerted by the Milky Way (as discussed in Section~\ref{ssec: tangential}), the escape velocity would be slightly superior
compared to such gravitational recoil kicks.\footnote{If the original M4 progenitor lost at least 60\% of its initial mass, similar structural parameters would still yield $v_{\rm esc} > 35 \, \kms$, while a mass loss of at least 80\% yields $v_{\rm esc} > 49 \, \kms$, preventing escape from non-spinning merging black holes of mass ratio greater than 12.}

To conclude, current models of dynamical evolution suggest that the formation and retention of an IMBH in M4 is feasible, although not necessarily likely. A diffuse population of stellar remnants would, on physical grounds, be a much more natural explanation of the excess mass we found in the core of M4.

\begin{figure}
\centering
\includegraphics[width=0.98\hsize]{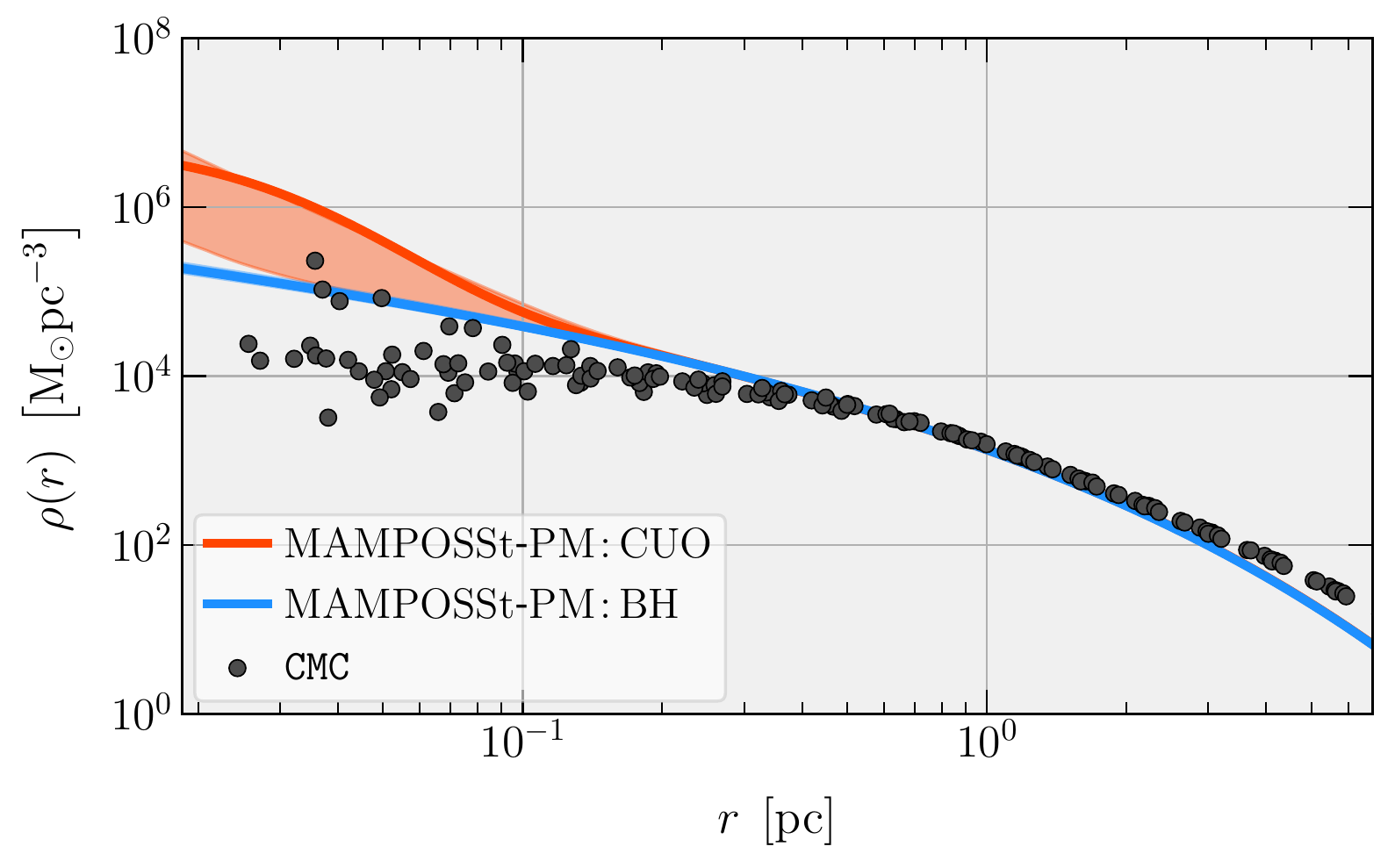}
\caption{\textit{Comparison of density profiles}: Comparison of the total mass density profiles of M4 (i.e., accounting for all remnants and luminous counterparts), estimated by \mpo\ (with the high-$\mu$ star discarded) and by \cmc\ (\emph{black circles}, gathered within $\sim 700$~Myr around our best snapshot, to better visualise the uncertainties on the simulaiton). The \emph{blue} line and its shaded region represent best likelihood and 16th-84th percentile region of the \mpo\ fit of a model assuming a central black hole alone (model~26), while the \emph{orange} counterparts relate to the model with a cluster of unseen objects (CUO, model~27).
The skewed marginal \mpo\ mass distribution of the CUO in M4 leads to a higher mode than the median, much in line with the maximum likelihood. This comparison shows that although a population of stellar remnants is a possible explanation for our fits, it should be much denser than what we are able to reproduce with our Monte-Carlo dynamical simulations. Thus, the existence of an IMBH cannot be ruled out, as the \cmc\ model agrees better (uncertainties accounted) to the respective \mpo\ fit.
}
\label{fig: bh-population}
\end{figure}

\subsection{A compact stellar-mass black hole population?}

As mentioned above, from the theoretical side, it would be a more natural explanation if our fits could be explained by a population of segregated stellar remnants such as 
proposed by \cite{Mann+19}, \cite{Zocchi+19}, \cite{Vitral&Mamon21}, and Paper~I.
Indeed, when removing the high-$\mu$ star, our fits yield a $932 \, \msun$ mass excess of 3D half-mass radius equal to 0.04~pc. We test the feasibility of this result, looking for M4 analogues in our \cmc\ models, similarly as what we did in Paper~I. 
The \cmc\ model that best matched the density profile of our fits presented a remaining segregated black hole population of $64 \, \msun$. Hence, our \mpo\ fits to the kinematic data without the high-$\mu$ star predict a black hole population more than ten times more massive than our best \cmc\ model (Figure~\ref{fig: bh-population} displays a comparison of mass density profiles). These numbers do not change significantly when accounting for other compact objects (i.e., white dwarfs and neutron stars) since they remain much less segregated than the black holes, and tend to mix within the stellar component of the cluster.

In fact, the retention of black holes in a GC is mediated primarily by its initial scale virial radius \citep{Kremer+20}. If a cluster starts with a low virial radius (i.e., denser), the rate of dynamical interactions will be faster, and its original black hole population will be quickly ejected through binary-mediated encounters. When this happens, black hole binary burning \citep{Kremer+20p} will no longer be effective and the cluster's core will collapse, forming a characteristic core-collapse structure. On the other hand, if a cluster starts with high virial radius, its original black hole population will take much longer to be ejected, and other luminous components will not be able to effectively populate the inner regions, forming a cored inner density profile. M4 seems to fall right in the midway scenario, where it has not yet reached core-collapse (as our fitted tangential velocity anisotropy suggests), but lost nonetheless an important fraction of its initial black hole population, thus departing from an usual cored density profile (see Figure~\ref{fig: cc-comp} for a comparison of the surface brightness profiles of between core-collapsed NGC~6397, cored NGC~3201, and M4).

One possibility to probe this difference between the \cmc\ models and our \mpo\ fits is to take into account the past dynamical history of M4. As argued in Section~\ref{ssec: tangential}, we expect the mean stellar mass of M4 to be higher, given its more intense tidal interactions \citep{Bianchini+17}. 
If the Milky Way tides are sufficiently strong to strip out part of the black hole population as it scatters to outer regions,\footnote{This can happen after three-body interactions in a typical black hole--black hole binary burning scenario \citep{Kremer+20p}.} one would also expect that the mean mass of the remaining black holes will be increased to higher masses compared to the mean mass of $13 \, \msun$ obtained with our best \cmc\ match.
It is however speculative that the trend proposed by \cite{Bianchini+17} extends to higher stellar masses, and it is more likely that the increase of mean stellar mass is due primarily to the escape of much lower mass stars, rather than low mass stellar-mass black holes. 
Another possibility would then be that the best \mpo\ model is the one with an IMBH of $792^{+253}_{-217} \, \msun$ ($820^{+186}_{-307} \, \msun$) when adding (removing) the high-$\mu$ star, in which case the density profile of our best \cmc\ match agrees better with the one predicted by \mpo\ (see Figure~\ref{fig: bh-population}). However, the simulation itself is not able to form such a massive black hole either.

In summary, although our \mpo\ fits could in principle be associated with a population of stellar-mass black holes in M4, this population should be much more numerous than what we can reproduce from idealised dynamical simulations. For this reason, we are not able to discard the possibility that M4 could have a low mass IMBH, although we could not reproduce its formation with the simulations either.
Whatever the detailed mechanism, our fits of M4 suggest an exotic mass excess, formed either by a low mass IMBH or by a super compact black hole population.

\section{Summary \& conclusions} \label{sec: conclusion}

We performed mass-anisotropy Jeans modelling of globular cluster M4 with the Bayesian code \mpo, following the same prescriptions from Paper~I, with data from \hst\ and with the \gaia\ EDR3 proper motion catalogue. We fit structural parameters such as mass, scale radius and S\'ersic index of the density profile, much in agreement with previous estimates \citep[e.g.,][]{Kimmig+15,Baumgardt&Hilker18} and more importantly, we simultaneously model the velocity anisotropy and inner mass excess in this cluster.

Similar to \cite{Vasiliev&Baumgardt&Baumgardt21}, we find an inner isotropic profile and outer tangential anisotropy for M4, which we associate with this cluster having not yet reached core-collapse \citep{Baumgardt&Makino&Makino03}, as well as with intense past tidal interactions \citep{Bianchini+17}. Its intense tidal interactions suggest that M4's progenitor was much more massive, and that its mean stellar mass is elevated due to the escape of low mass stars on radial orbits.

Our fits of an inner mass excess are the first to suggest an inner dark mass of roughly $800 \, \msun$ in M4.\footnote{The \mpo\ fits yield $7.9^{+2.5}_{-2.2} \times 10^{2} \, \msun$ for an IMBH scenario (i.e., model~6 from Table~\ref{tab: results-full}), and $9.4^{+1.7}_{-3.3} \times 10^{2} \, \msun$ for an CUO scenario (i.e., model~7 from Table~\ref{tab: results-full}).} We also fitted the extension of such a mass and find a relatively small scale radius (namely, 0.016~pc), which although not point-like as expected for an IMBH, it still remains too small to be reliably associated with a collection of dark remnants.
This small extent is associated with a single high velocity star that we label the \textit{high-$\mu$ star}, whose proper motion fit seems robust within its uncertainties, but extra checks reveal a bright nearby neighbour $0\farcs23$ away that undermines several photometric measurements, and we thus decided to remove it.  After doing so, we still find a $\sim 800 \, \msun$ mass excess,\footnote{The \mpo\ fits yield $8.2^{+1.9}_{-3.1}  \times 10^{2} \, \msun$ for an IMBH scenario (i.e., model~26 from Table~\ref{tab: results-full}), and $9.3^{+2.2}_{-3.6} \times 10^{2} \, \msun$ for a CUO scenario (i.e., model~27 from Table~\ref{tab: results-full}).} but now having a twice larger scale radius, which approaches better an extended concentration of unresolved stellar remnants as found in Paper~I. Nevertheless, it is important to mention that the removed star has a proper motion vector consistent with the scenario where it is coming from the cluster's centre, and our analyses with mock data reveal that in case M4 has a central IMBH, there is 33\% of chance that one high-$\mu$ star is measured in the cluster's inner regions.

Next, we used dynamical Monte Carlo $N$-body models constructed with the \cmc\ code to test whether the extended dark population that was found when removing the high-$\mu$ star is a viable solution. Our best \cmc\ model matching the surface brightness and velocity dispersion profile of M4 involves a cluster with a remaining black hole population of $64 \, \msun$, composed of five segregated black holes. Our fits suggest, however, that the concentration of stellar remnants in M4 is more than ten times more massive than what we could reproduce with the \cmc\ models, likely pointing to a super compact population of massive stellar-mass black holes, whose feasibility is also debatable.

One might then wonder whether an IMBH could be masquerading as a super compact population of remnants. The formation of an IMBH with a mass similar to what we fit is feasible through early runaway merger scenarios such as proposed in \cite{Gonzalez+21}. Its retention could be possible if the cluster was more massive in the past, as we expect from the strong tidal stripping of the Milky Way given M4's very small current pericentre and also from our velocity anisotropy fits. In such a case, the IMBH eventually merges with other segregated black holes and receives gravitational recoil kicks, whose velocities should not overcome M4's escape velocity given the high mass ratios between the putative IMBH and other black holes, namely $M_{\rm IMBH} / M_{\rm BH} \sim 20$. 
Even though such a mechanism is feasible, it should be taken with caution, given the uncertain assumptions involved in runaway merging models.

In essence, the dark central mass in M4 is likely an exotic scenario, whether composed by a central IMBH, or by a super compact black hole population. Although we find it hard to fully explain either scenario from the theoretical side, we highlight that a concentration of stellar remnants would still seem like a more realistic solution, from recent studies of globular clusters (e.g. \citealt{Kremer+20}; \citealt{Gieles+21}; Paper~I). 

The physics of the inner mass of M4 deserves additional study. It will be worthwhile improving our analysis with the next data release from \gaia. We propose follow-up observations of M4 from more imaging using \hst, and the {\sl James Webb Space Telescope}, which could increase the proper motion baseline of this cluster, increasing the number of well-measured tracers, as well as provide more accurate photometry for the high-$\mu$ star. Our results hence set an interesting target for future observational campaigns and open an important debate concerning the closest globular cluster to our Sun. 


\section*{Acknowledgements}

We thank the anonymous referee for the constructive report, with comments that have helped us to improve the quality of our results and clarify some descriptions in the manuscript.
We also acknowledge Roeland van der Marel and the HSTPROMO collaboration\footnote{\url{https://www.stsci.edu/~marel/hstpromo.html}.} for useful comments.
\\
Eduardo Vitral was funded by an AMX doctoral grant from \'Ecole Polytechnique.
Kyle Kremer is supported by an NSF Astronomy and Astrophysics Postdoctoral Fellowship under award AST-2001751. 
LRB acknowledges support by MIUR under PRIN programme \#2017Z2HSMF and by PRIN-INAF 2012 and 2019.
\\
Support for this work was provided by a grant for \hst\ program 13297 provided by the Space Telescope Science Institute, which is operated by AURA, Inc., under NASA contract NAS 5-26555. This work has made use of data from the European Space Agency (ESA) mission \gaia\ (\url{https://www.cosmos.esa.int/gaia}), processed by the \gaia\ Data Processing and Analysis Consortium (DPAC, \url{https://www.cosmos.esa.int/web/gaia/dpac/consortium}). Funding for the DPAC has been provided by national institutions, in particular the institutions participating in the \gaia\ Multilateral Agreement.
We greatly benefited from the public software {\sc Python} \citep{VanRossum09} packages 
{\sc BALRoGO} \citep{Vitral21},
{\sc Scipy} \citep{Jones+01},
{\sc Numpy} \citep{vanderWalt11} and
{\sc Matplotlib} \citep{Hunter07}. We also used the {\sc Spyder} Integrated Development Environment \citep{raybaut2009spyder}.

\section*{Data Availability}

The data that support the plots within this paper and other findings of this study are available from the corresponding author upon reasonable request.



\bibliographystyle{mnras}
\bibliography{src} 




\newpage

\appendix

\section{Extra material}

\begin{table*}
\caption{Main results of the \mpo\ mass-modelling fit of NGC~6121.}
\label{tab: results-full}
\centering
\renewcommand{\arraystretch}{1.7}
\tabcolsep=2.7pt
\footnotesize
\begin{tabular}{ll@{\hspace{2mm}}ccrrrrrrrrr}
\hline\hline             
\multicolumn{1}{c}{Model} &
\multicolumn{1}{c}{Cluster ID} &
\multicolumn{1}{c}{Test} &
\multicolumn{1}{c}{$R^{-1}$} & 
\multicolumn{1}{c}{$\beta_{0}$} &
\multicolumn{1}{c}{$\beta_{\rm out}$} &
\multicolumn{1}{c}{$r_{\rm GC}$} &
\multicolumn{1}{c}{$n_{\rm GC}$} &
\multicolumn{1}{c}{$M_{\rm GC}$} &
\multicolumn{1}{c}{$r_{\rm CUO}$} &
\multicolumn{1}{c}{$M_{\rm CUO}$} &
\multicolumn{1}{c}{$M_{\rm BH}$} &
\multicolumn{1}{c}{$\Delta \rm AICc$} \\
\multicolumn{1}{c}{} &
\multicolumn{1}{c}{} &
\multicolumn{1}{c}{} &
\multicolumn{1}{c}{} &
\multicolumn{1}{c}{} &
\multicolumn{1}{c}{} &
\multicolumn{1}{c}{[pc]} & 
\multicolumn{1}{c}{} & 
\multicolumn{1}{c}{[$10^{4} $ M$_\odot$]} & 
\multicolumn{1}{c}{[pc]} & 
\multicolumn{1}{c}{[M$_\odot$]} &
\multicolumn{1}{c}{[M$_\odot$]} &
\multicolumn{1}{c}{} \\ 
\multicolumn{1}{c}{(1)} &
\multicolumn{1}{c}{(2)} &
\multicolumn{1}{c}{(3)} &
\multicolumn{1}{c}{(4)} &
\multicolumn{1}{c}{(5)} & 
\multicolumn{1}{c}{(6)} & 
\multicolumn{1}{c}{(7)} & 
\multicolumn{1}{c}{(8)} & 
\multicolumn{1}{c}{(9)} & 
\multicolumn{1}{c}{(10)} & 
\multicolumn{1}{c}{(11)} &
\multicolumn{1}{c}{(12)} & 
\multicolumn{1}{c}{(13)} \\ 
\hline
1 & NGC~6121 & \multicolumn{1}{c}{--} & $  0.001 $ & \multicolumn{1}{c}{\darkg{\bf 0 }} & \multicolumn{1}{c}{\darkg{\bf 0 }} & $  2.164^{+ 0.057}_{- 0.069} $ & $  2.14^{+ 0.07}_{- 0.05} $ & $  8.51^{+ 0.18}_{- 0.17} $ & \multicolumn{1}{c}{--} & \multicolumn{1}{c}{--} & \multicolumn{1}{c}{--} &  62.47 \\
2 & NGC~6121 & \multicolumn{1}{c}{--} & $  0.004 $ & \multicolumn{1}{c}{\darkg{\bf 0 }} & \multicolumn{1}{c}{\darkg{\bf 0 }} & $  2.348^{+ 0.087}_{- 0.080} $ & $  2.09^{+ 0.07}_{- 0.05} $ & $  8.54^{+ 0.23}_{- 0.14} $ & \multicolumn{1}{c}{--} & \multicolumn{1}{c}{--} & $  927^{+ 140}_{- 278} $ &  30.55 \\
3 & NGC~6121 & \multicolumn{1}{c}{--} & $  0.003 $ & \multicolumn{1}{c}{\darkg{\bf 0 }} & \multicolumn{1}{c}{\darkg{\bf 0 }} & $  2.366^{+ 0.079}_{- 0.090} $ & $  2.08^{+ 0.08}_{- 0.04} $ & $  8.60^{+ 0.17}_{- 0.20} $ & $  0.014^{+ 0.006}_{- 0.013} $ & $  910^{+ 216}_{- 226} $ & \multicolumn{1}{c}{--} &  32.50 \\
4 & NGC~6121 & \multicolumn{1}{c}{--} & $  0.015 $ & \multicolumn{1}{c}{\darkg{\bf 0 }} & \multicolumn{1}{c}{\darkg{\bf 0 }} & $  2.356^{+ 0.092}_{- 0.077} $ & $  2.10^{+ 0.06}_{- 0.07} $ & $  8.56^{+ 0.21}_{- 0.16} $ & $  0.017^{+ 0.087}_{- 0.016} $ & $  202^{+ 636}_{- 173} $ & $  729^{+ 178}_{- 671} $ &  34.52 \\
5 & NGC~6121 & $\beta(r)$ & $  0.002 $ & $  0.18^{+ 0.03}_{- 0.07} $ & $ -0.41^{+ 0.08}_{- 0.05} $ & $  2.056^{+ 0.069}_{- 0.059} $ & $  2.12^{+ 0.08}_{- 0.04} $ & $  8.23^{+ 0.19}_{- 0.16} $ & \multicolumn{1}{c}{--} & \multicolumn{1}{c}{--} & \multicolumn{1}{c}{--} &  23.40 \\
\rowcolor{lavender} 6 & NGC~6121 & $\beta(r)$ & $  0.007 $ & $  0.04^{+ 0.07}_{- 0.05} $ & $ -0.36^{+ 0.06}_{- 0.08} $ & $  2.207^{+ 0.095}_{- 0.070} $ & $  2.08^{+ 0.09}_{- 0.04} $ & $  8.30^{+ 0.18}_{- 0.20} $ & \multicolumn{1}{c}{--} & \multicolumn{1}{c}{--} & $  792^{+ 253}_{- 217} $ &  0.00 \\
\rowcolor{lavender} 7 & NGC~6121 & $\beta(r)$ & $  0.008 $ & $  0.06^{+ 0.05}_{- 0.07} $ & $ -0.37^{+ 0.08}_{- 0.06} $ & $  2.218^{+ 0.094}_{- 0.073} $ & $  2.10^{+ 0.06}_{- 0.06} $ & $  8.21^{+ 0.26}_{- 0.11} $ & $  0.016^{+ 0.005}_{- 0.015} $ & $  939^{+ 166}_{- 331} $ & \multicolumn{1}{c}{--} &  1.80 \\
8 & NGC~6121 & $\beta(r)$ & $  0.009 $ & $  0.04^{+ 0.06}_{- 0.05} $ & $ -0.34^{+ 0.05}_{- 0.10} $ & $  2.232^{+ 0.084}_{- 0.086} $ & $  2.10^{+ 0.07}_{- 0.06} $ & $  8.27^{+ 0.21}_{- 0.17} $ & $  0.004^{+ 0.127}_{- 0.002} $ & $ \darkor{ 15^{+ 793}_{- 0}} $ & $  847^{+ 10}_{- 776} $ &  3.87 \\
9 & NGC~6121 & $(\alpha_{0}, \, \delta_{0})$ & $  0.002 $ & $  0.16^{+ 0.05}_{- 0.05} $ & $ -0.39^{+ 0.06}_{- 0.08} $ & $  2.045^{+ 0.078}_{- 0.049} $ & $  2.12^{+ 0.08}_{- 0.05} $ & $  8.21^{+ 0.21}_{- 0.14} $ & \multicolumn{1}{c}{--} & \multicolumn{1}{c}{--} & \multicolumn{1}{c}{--} &  122850.31 \\
10 & NGC~6121 & $(\alpha_{0}, \, \delta_{0})$ & $  0.005 $ & $  0.05^{+ 0.06}_{- 0.06} $ & $ -0.38^{+ 0.08}_{- 0.06} $ & $  2.230^{+ 0.072}_{- 0.092} $ & $  2.11^{+ 0.05}_{- 0.07} $ & $  8.27^{+ 0.20}_{- 0.17} $ & \multicolumn{1}{c}{--} & \multicolumn{1}{c}{--} & $  848^{+ 199}_{- 264} $ &  122826.98 \\
11 & NGC~6121 & $(\alpha_{0}, \, \delta_{0})$ & $  0.005 $ & $  0.01^{+ 0.09}_{- 0.03} $ & $ -0.36^{+ 0.07}_{- 0.08} $ & $  2.225^{+ 0.089}_{- 0.079} $ & $  2.09^{+ 0.07}_{- 0.05} $ & $  8.24^{+ 0.23}_{- 0.15} $ & $  0.017^{+ 0.005}_{- 0.015} $ & $  942^{+ 175}_{- 323} $ & \multicolumn{1}{c}{--} &  122829.06 \\
12 & NGC~6121 & $(\alpha_{0}, \, \delta_{0})$ & $  0.020 $ & $  0.04^{+ 0.06}_{- 0.06} $ & $ -0.34^{+ 0.05}_{- 0.09} $ & $  2.244^{+ 0.070}_{- 0.098} $ & $  2.11^{+ 0.06}_{- 0.07} $ & $  8.34^{+ 0.12}_{- 0.24} $ & $  0.009^{+ 0.123}_{- 0.007} $ & $  778^{+ 56}_{- 743} $ & $ \darkor{ 28^{+ 830}_{- 0}} $ &  122830.95 \\
13 & NGC~6121 & $\sigma_{\mu}$ & $  0.005 $ & $  0.18^{+ 0.06}_{- 0.04} $ & $ -0.47^{+ 0.11}_{- 0.10} $ & $  1.969^{+ 0.084}_{- 0.052} $ & $  2.15^{+ 0.08}_{- 0.06} $ & $  7.91^{+ 0.25}_{- 0.18} $ & \multicolumn{1}{c}{--} & \multicolumn{1}{c}{--} & \multicolumn{1}{c}{--} &  80807.35 \\
14 & NGC~6121 & $\sigma_{\mu}$ & $  0.007 $ & $  0.07^{+ 0.06}_{- 0.06} $ & $ -0.45^{+ 0.14}_{- 0.08} $ & $  2.142^{+ 0.096}_{- 0.090} $ & $  2.09^{+ 0.09}_{- 0.05} $ & $  7.93^{+ 0.24}_{- 0.24} $ & \multicolumn{1}{c}{--} & \multicolumn{1}{c}{--} & $  876^{+ 230}_{- 250} $ &  80781.82 \\
15 & NGC~6121 & $\sigma_{\mu}$ & $  0.006 $ & $  0.04^{+ 0.09}_{- 0.04} $ & $ -0.39^{+ 0.08}_{- 0.14} $ & $  2.155^{+ 0.092}_{- 0.097} $ & $  2.09^{+ 0.09}_{- 0.05} $ & $  7.92^{+ 0.25}_{- 0.24} $ & $  0.013^{+ 0.011}_{- 0.012} $ & $  1007^{+ 186}_{- 329} $ & \multicolumn{1}{c}{--} &  80783.78 \\
16 & NGC~6121 & $\sigma_{\mu}$ & $  0.015 $ & $  0.02^{+ 0.11}_{- 0.02} $ & $ -0.38^{+ 0.07}_{- 0.15} $ & $  2.169^{+ 0.084}_{- 0.104} $ & $  2.12^{+ 0.06}_{- 0.09} $ & $  7.89^{+ 0.28}_{- 0.20} $ & $  0.030^{+ 0.106}_{- 0.028} $ & $  556^{+ 354}_{- 519} $ & $  466^{+ 446}_{- 407} $ &  80785.83 \\
17 & NGC~6121 & Bulk $\mu$ & $  0.006 $ & $  0.15^{+ 0.06}_{- 0.04} $ & $ -0.38^{+ 0.05}_{- 0.08} $ & $  2.073^{+ 0.050}_{- 0.078} $ & $  2.15^{+ 0.05}_{- 0.08} $ & $  8.26^{+ 0.16}_{- 0.19} $ & \multicolumn{1}{c}{--} & \multicolumn{1}{c}{--} & \multicolumn{1}{c}{--} &  122841.25 \\
18 & NGC~6121 & Bulk $\mu$ & $  0.007 $ & $  0.07^{+ 0.04}_{- 0.07} $ & $ -0.39^{+ 0.10}_{- 0.05} $ & $  2.222^{+ 0.079}_{- 0.085} $ & $  2.10^{+ 0.06}_{- 0.06} $ & $  8.26^{+ 0.21}_{- 0.16} $ & \multicolumn{1}{c}{--} & \multicolumn{1}{c}{--} & $  872^{+ 166}_{- 298} $ &  122817.86 \\
19 & NGC~6121 & Bulk $\mu$ & $  0.009 $ & $  0.04^{+ 0.07}_{- 0.05} $ & $ -0.36^{+ 0.07}_{- 0.08} $ & $  2.217^{+ 0.092}_{- 0.075} $ & $  2.12^{+ 0.05}_{- 0.08} $ & $  8.27^{+ 0.20}_{- 0.18} $ & $  0.009^{+ 0.011}_{- 0.008} $ & $  855^{+ 243}_{- 249} $ & \multicolumn{1}{c}{--} &  122819.54 \\
20 & NGC~6121 & Bulk $\mu$ & $  0.009 $ & $  0.02^{+ 0.08}_{- 0.04} $ & $ -0.35^{+ 0.05}_{- 0.09} $ & $  2.269^{+ 0.040}_{- 0.125} $ & $  2.12^{+ 0.05}_{- 0.08} $ & $  8.34^{+ 0.12}_{- 0.25} $ & $  0.023^{+ 0.118}_{- 0.022} $ & $ \darkor{ 802^{+ 0}_{- 770}} $ & $  178^{+ 678}_{- 114} $ &  122821.61 \\
21 & NGC~6121 & $\epsilon_{\mathrm{sys}}$ & $  0.002 $ & $  0.17^{+ 0.04}_{- 0.05} $ & $ -0.40^{+ 0.07}_{- 0.06} $ & $  2.032^{+ 0.068}_{- 0.060} $ & $  2.13^{+ 0.08}_{- 0.05} $ & $  8.10^{+ 0.20}_{- 0.14} $ & \multicolumn{1}{c}{--} & \multicolumn{1}{c}{--} & \multicolumn{1}{c}{--} &  122848.59 \\
22 & NGC~6121 & $\epsilon_{\mathrm{sys}}$ & $  0.010 $ & $  0.03^{+ 0.07}_{- 0.04} $ & $ -0.36^{+ 0.07}_{- 0.08} $ & $  2.207^{+ 0.072}_{- 0.092} $ & $  2.12^{+ 0.05}_{- 0.08} $ & $  8.15^{+ 0.20}_{- 0.17} $ & \multicolumn{1}{c}{--} & \multicolumn{1}{c}{--} & $  850^{+ 218}_{- 258} $ &  122824.48 \\
23 & NGC~6121 & $\epsilon_{\mathrm{sys}}$ & $  0.007 $ & $  0.04^{+ 0.06}_{- 0.06} $ & $ -0.34^{+ 0.05}_{- 0.09} $ & $  2.196^{+ 0.095}_{- 0.074} $ & $  2.10^{+ 0.07}_{- 0.05} $ & $  8.12^{+ 0.22}_{- 0.15} $ & $ \darkor{ 0.001^{+ 0.021}_{- 0.000}} $ & $  912^{+ 222}_{- 284} $ & \multicolumn{1}{c}{--} &  122826.48 \\
24 & NGC~6121 & $\epsilon_{\mathrm{sys}}$ & $  0.015 $ & $  0.05^{+ 0.05}_{- 0.06} $ & $ -0.35^{+ 0.06}_{- 0.08} $ & $  2.233^{+ 0.059}_{- 0.108} $ & $  2.12^{+ 0.05}_{- 0.08} $ & $  8.17^{+ 0.17}_{- 0.20} $ & $  0.026^{+ 0.107}_{- 0.024} $ & $  652^{+ 209}_{- 618} $ & $  287^{+ 583}_{- 222} $ &  122828.25 \\
25 & NGC~6121 & $v_{\star}$ & $  0.002 $ & $  0.16^{+ 0.04}_{- 0.05} $ & $ -0.41^{+ 0.08}_{- 0.06} $ & $  2.036^{+ 0.094}_{- 0.035} $ & $  2.14^{+ 0.06}_{- 0.07} $ & $  8.19^{+ 0.25}_{- 0.11} $ & \multicolumn{1}{c}{--} & \multicolumn{1}{c}{--} & \multicolumn{1}{c}{--} &  122824.23 \\
\rowcolor{lavender} 26 & NGC~6121 & $v_{\star}$ & $  0.004 $ & $  0.04^{+ 0.07}_{- 0.05} $ & $ -0.39^{+ 0.09}_{- 0.05} $ & $  2.233^{+ 0.060}_{- 0.106} $ & $  2.11^{+ 0.06}_{- 0.07} $ & $  8.29^{+ 0.18}_{- 0.19} $ & \multicolumn{1}{c}{--} & \multicolumn{1}{c}{--} & $  820^{+ 186}_{- 307} $ &  122808.88 \\
\rowcolor{lavender} 27 & NGC~6121 & $v_{\star}$ & $  0.017 $ & $  0.03^{+ 0.08}_{- 0.04} $ & $ -0.34^{+ 0.04}_{- 0.10} $ & $  2.240^{+ 0.074}_{- 0.099} $ & $  2.09^{+ 0.08}_{- 0.05} $ & $  8.32^{+ 0.16}_{- 0.22} $ & $  0.034^{+ 0.021}_{- 0.032} $ & $  932^{+ 220}_{- 355} $ & \multicolumn{1}{c}{--} &  122809.26 \\
28 & NGC~6121 & $v_{\star}$ & $  0.013 $ & $  0.06^{+ 0.05}_{- 0.07} $ & $ -0.37^{+ 0.08}_{- 0.06} $ & $  2.197^{+ 0.116}_{- 0.055} $ & $  2.11^{+ 0.05}_{- 0.07} $ & $  8.16^{+ 0.31}_{- 0.06} $ & $  0.041^{+ 0.079}_{- 0.039} $ & $ \darkor{ 936^{+ 0}_{- 895}} $ & $ \darkor{ 17^{+ 764}_{- 0}} $ &  122811.91 \\
\hline
\end{tabular}
\parbox{\hsize}{\textit{Notes}: Columns are 
\textbf{(1)} Model number; 
\textbf{(2)} Cluster ID; 
\textbf{(3)} Test type: "$\beta(r)$" for a free anisotropy model, ``$(\alpha_{0}, \, \delta_{0})$'' for the test of a different centre (\protect\citealt{Goldsbury+10}), ``$\sigma_{\mu}$'' for the test with half of the standard error threshold, ``Bulk $\mu$'' for the test setting the \hst\ bulk proper motion as the one from \protect\cite{Vasiliev&Baumgardt&Baumgardt21}, ``$\epsilon_{\rm sys}$'' for tests considering the \gaia\ systematics according to eq.~3 from \protect\cite{Vasiliev&Baumgardt&Baumgardt21} and ``$v_{\star}$'' for the subset without the high velocity star (Section~\ref{ssec: high-v}); 
\textbf{(4)} MCMC convergence criterion ($R^{-1} \leq 0.02$ is considered as properly converged);
\textbf{(5)} anisotropy value at $r=0$;
\textbf{(6)} anisotropy value at the data's most distant projected radius (i.e., 12.6 arcmin);
\textbf{(7)} S\'ersic projected half mass radius $R_{\rm e}$ (in pc) of the mass density profile of the globular cluster;
\textbf{(8)} S\'ersic index $n$ of the mass density profile of the globular cluster;
\textbf{(9)} Total globular cluster mass (without dark central component), in M$_{\odot}$;
\textbf{(10)} Plummer projected half mass radius $a_{\rm P}$ (in pc) of the mass density profile of the central sub-cluster of unresolved objects (CUO);
\textbf{(11)} Total mass of the CUO, in M$_{\odot}$;
\textbf{(12)} Central black hole mass, in M$_{\odot}$;
\textbf{(13)} Difference in AICc (eq.~[14]) relative to model 6.
We highlight the maximum likelihood values in \emph{orange} when they were outside the 16-84 percentiles of the posterior distribution.
The uncertainties are respective to the 16th and 84th percentiles of the marginal distributions.
The lines coloured in \textit{lavender} indicate our preferred models, i.e., models with a central black hole (6 and 26) or with a central CUO (7 and 27), for both the cases with (6--7) and without (26--27) the high velocity star. We did not consider the AICc diagnosis when the dataset was different from the respective standard model (i.e., the one used in models 1--8).}
\end{table*}


\begin{table}
\caption{Observed velocity dispersion depicted in Figure~\ref{fig: vdisp}.}
\label{tab: vdisp}
\centering
\renewcommand{\arraystretch}{1.4}
\tabcolsep=3.5pt
\footnotesize
\begin{tabular}{rrcc}
\hline\hline   
\multicolumn{1}{c}{$R_{\rm proj}$} &
\multicolumn{1}{c}{$\epsilon_{R_{\rm proj}}$} &
\multicolumn{1}{c}{$\sigma_{\mu}$} &
\multicolumn{1}{c}{$\epsilon_{\sigma_{\mu}}$} \\
\multicolumn{1}{c}{[arcsec]} &
\multicolumn{1}{c}{[arcsec]} &
\multicolumn{1}{c}{[$\masyr$]} &
\multicolumn{1}{c}{[$\masyr$]} \\ 
\hline
      1.33  &     0.13  &  1.095  &  0.489 \\
      2.84  &     0.39  &  0.675  &  0.181 \\
      5.65  &     0.93  &  0.730  &  0.160 \\
     10.58  &     1.76  &  0.607  &  0.048 \\
     21.00  &     3.48  &  0.604  &  0.025 \\
     39.92  &     6.99  &  0.597  &  0.014 \\
     75.52  &   13.39   & 0.578   & 0.009 \\
    144.38  &    28.69  &  0.526  &  0.012 \\
    290.83  &    51.95  &  0.468  &  0.008 \\
    535.05  &    98.74  &  0.381  &  0.006 \\
\hline
\end{tabular}
\parbox{\hsize}{\textit{Notes}: 
 Columns are \textbf{(1)} Projected distance to the centre;
\textbf{(2)} Uncertainty on projected distance to the centre;
\textbf{(3)} Observed proper motion dispersion;
\textbf{(4)} Uncertainty on proper motion dispersion.
}
\end{table}

\begin{figure}
\centering
\includegraphics[width=0.95\hsize]{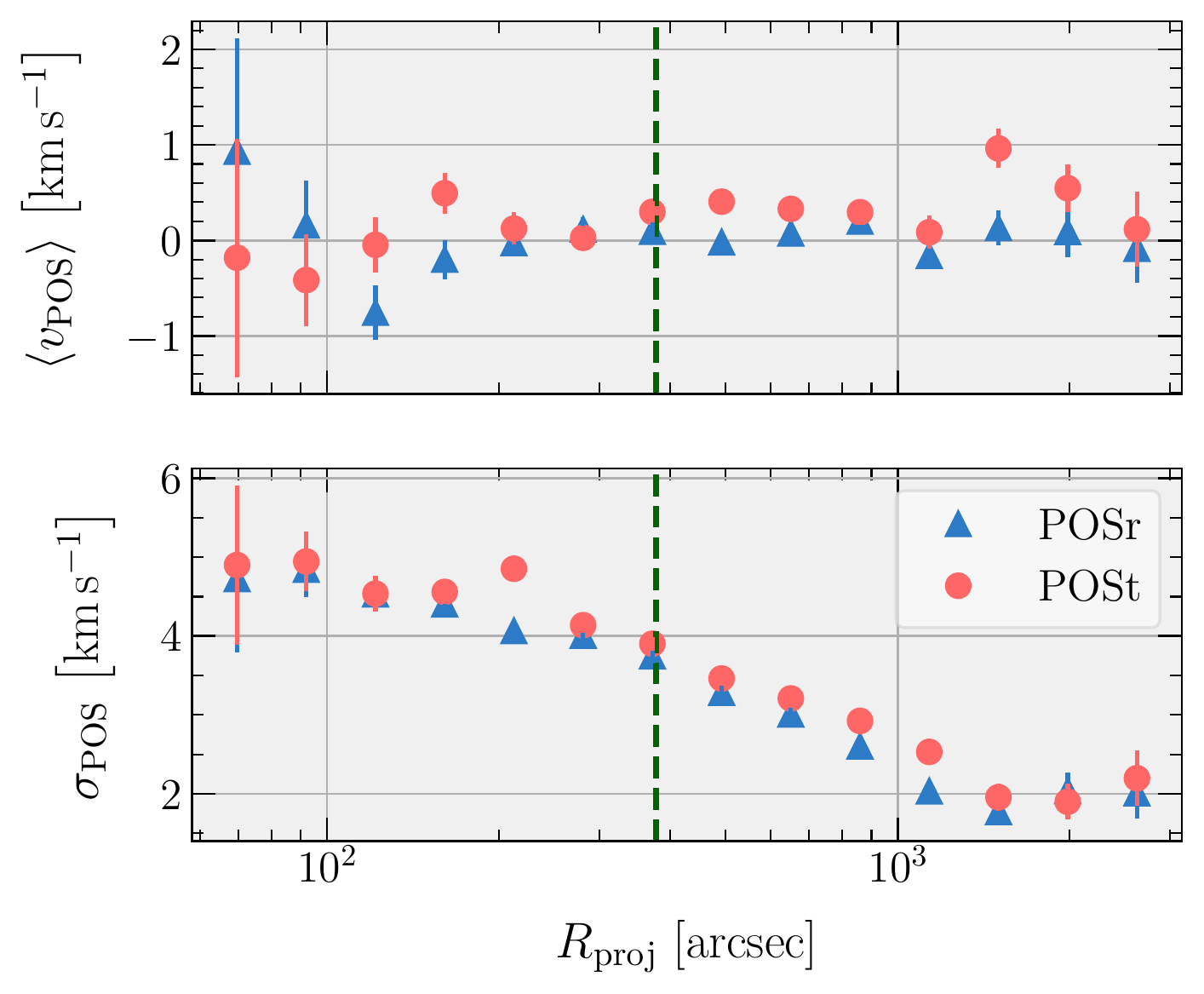}
\caption{
\textit{Plane of sky velocities:} Radial profiles of mean plane of sky velocity (top) and velocity
dispersion (bottom) of M4 from cleaned \gaia\ EDR3 data. The plane of sky motions are split between radial (POSr, blue triangles) and tangential (POSt, red circles) components. The dashed green vertical line displays the $2 \, R_{\rm e}$ limit we use in \mpo. Both the velocity dispersion and its mean (along with respective uncertainties) were calculated with the recipe from \citeauthor{vanderMarel&Anderson&Anderson10}~(\citeyear{vanderMarel&Anderson&Anderson10}, appendix~A).
This plots suggests that the analysed data does not suffer from significant tidal effects, which in turn seem to be visible on the velocity dispersion profile only beyond $10^3$ arcsec. The conversions from $\masyr$ to $\kms$ assumed the M4 distance from \protect\cite{Baumgardt&Vasiliev21}.
}
\label{fig: pos-m4}
\end{figure}

\begin{figure}
\centering
\includegraphics[width=0.95\hsize]{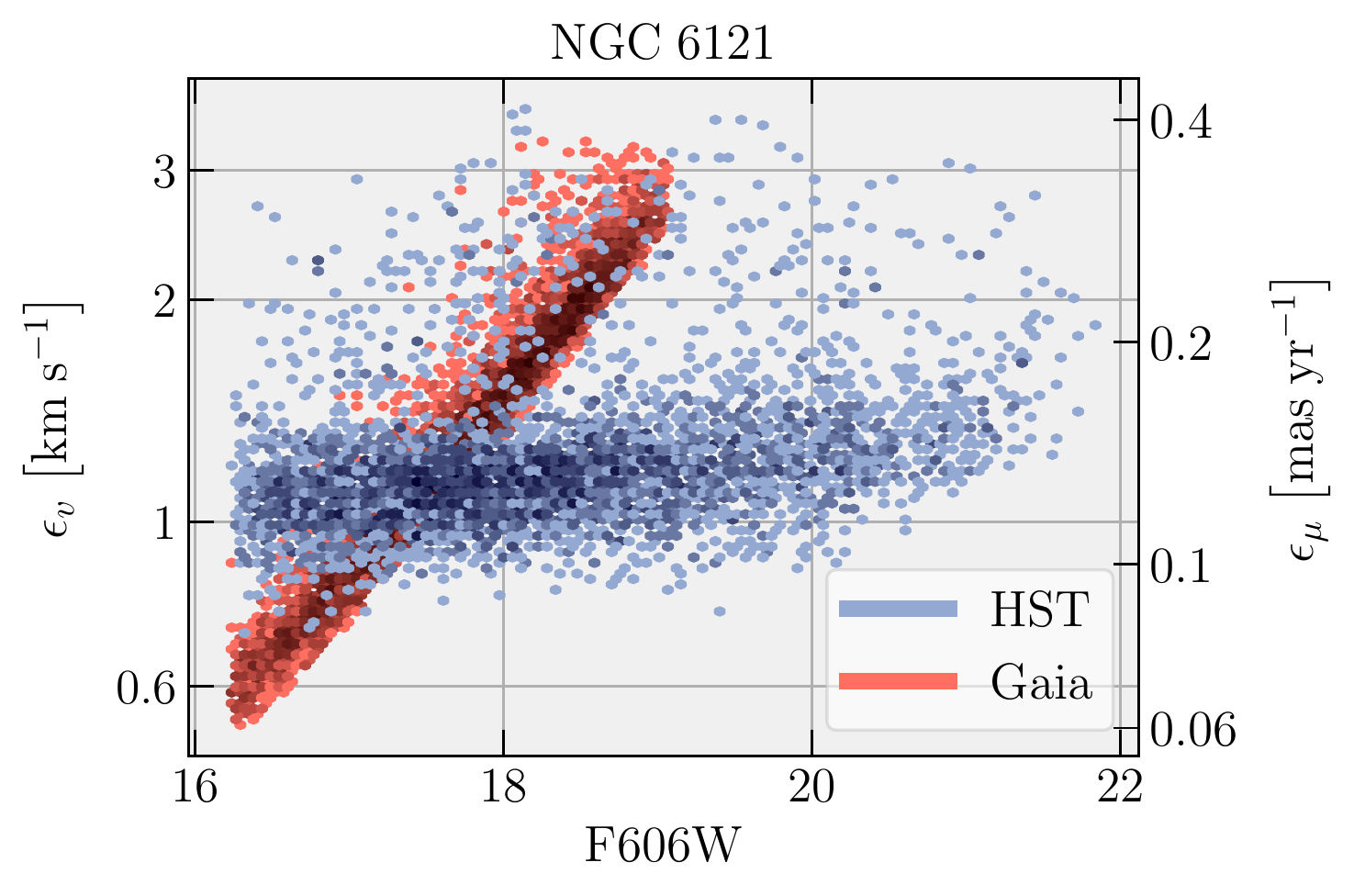}
\caption{
\textit{Error-magnitude relation:} The plot limits represent the respective limits of our cleaned data. \textbf{Blue} points relate to \textit{HST} data, while the \textbf{red} ones are from \textit{Gaia} EDR3 data. The conversions from $\masyr$ to $\kms$ assumed the M4 distance from \protect\cite{Baumgardt&Vasiliev21}, and errors are defined as in Paper~I. 
The expected effects of mass segregation on the velocity dispersion profile (see section~5.3.2 from Paper~I), calculated from the 1.15 ratio of 75th to 25th mass percentiles, translates to 0.043 $\masyr$ or 0.38 $\kms$. This effect is well below the typical proper motion uncertainties of our data.
}
\label{fig: err-mag-relation}
\end{figure}

\begin{figure}
\centering
\includegraphics[width=0.9\hsize]{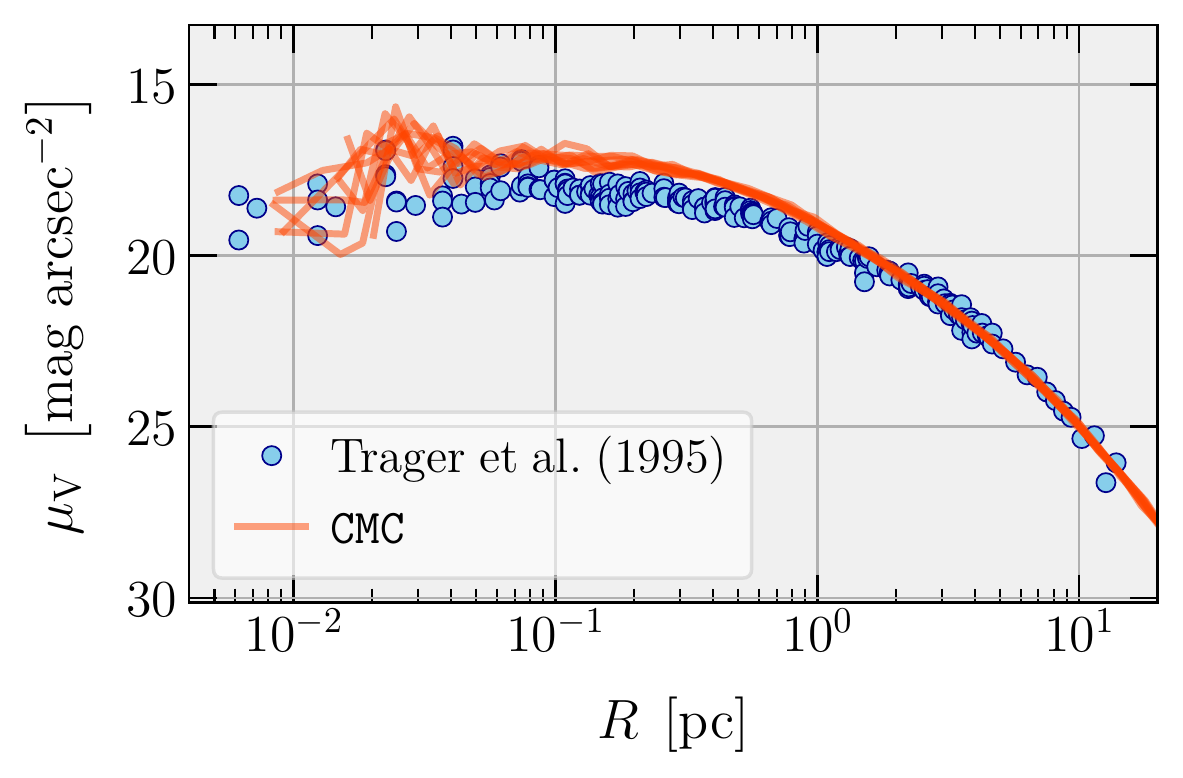}
\caption{
\textit{Match of surface brightness profile:} The plot depicts the agreement between the surface brightness profile of M4 observed by \protect\cite{Trager+95} as blue circles, and the ten closest snapshots from our best \cmc\ model as orange lines.
}
\label{fig: sbp-cmc}
\end{figure}



\bsp	
\label{lastpage}
\end{document}